\RequirePackage{fix-cm}
\documentclass[reqno]{amsproc}
\usepackage{amssymb,amsthm}

\usepackage{microtype}

\usepackage[citation-order,nobysame]{amsrefs}
\usepackage{xyzbib}

\usepackage{hyperref}

\usepackage{graphicx}
\usepackage{tikz}

\newtheorem{theorem}{Theorem}[section]
\newtheorem{proposition}[theorem]{Proposition}

\let\cite=\cites
\numberwithin{equation}{section}
\newcommand\wt{\widetilde}
\newcommand\eps{\varepsilon}

\newcommand\Rc{R_\mathrm{c}}
\newcommand\Qc{Q_\mathrm{c}}

\newcommand{\rme}{\mathrm{e}}
\newcommand{\rmi}{\mathrm{i}}
\newcommand{\rmd}{\mathrm{d}}

\begin{document}
\title{Arctic curve of the free-fermion 
six-vertex model \\ in an L-shaped domain}

\author{F. Colomo}
\address{INFN, Sezione di Firenze, 
Via G. Sansone 1, 50019 Sesto Fiorentino (FI), Italy}
\email{colomo@fi.infn.it}

\author{A. G. Pronko}
\address{Steklov Mathematical 
Institute,
Fontanka 27, 191023 St. Petersburg, Russia}
\email{agp@pdmi.ras.ru}

\author{A. Sportiello}
\address{LIPN, and CNRS, Universit\'e Paris 13, Sorbonne Paris Cit\'e,
99 Av. J.-B. Cl\'ement, 93430 Villetaneuse, France}
\email{Andrea.Sportiello@lipn.univ-paris13.fr}

\begin{abstract}
We consider the six-vertex model in an L-shaped domain of the
square lattice, with domain wall boundary conditions, in the case
of free-fermion vertex weights. We describe how the recently
developed `Tangent method' can be used to determine the form of the
arctic curve. The obtained result is in agreement with numerics.
\end{abstract}

\maketitle
\section{Introduction}

The thermodynamics of the six-vertex model with fixed boundary
conditions has attracted much attention in recent years, in
particular, as an example of a system exhibiting (in an appropriate
scaling limit) spatial phase separation phenomena
\cite{KZj-00,Zj-00,Zj-02,RP-06,CP-09,BL-13,RS-15,RS-16,ADSV-16,BCG-16,D-16,GBDJ-18}.
The model can be regarded as a nontrivial generalization of dimer
models, and in particular, under specific geometry and boundary
conditions, of the famous problem of domino tilings of the Aztec
diamond, where the celebrated Arctic Circle phenomenon was discovered
\cite{JPS-98}.

Among the many questions concerning this kind of effects, the shape of
the phase separating curves (known as arctic curves) is of prime
interest on its own right
\cite{CLP-98,CK-01,OR-01,FS-03,KO-06,KOS-06,KO-07,PS-05,PR-07,DfSg-14,Pet-14,RSn-16,BBCCR-17,BK-18,DfL-18,DfG-18},
and also in view of its relevance in relation with quantum quenches and
nonequilibrium transport in one-dimensional quantum spin chains
\cite{ADSV-16,S-17,CDlV-18}, and with spin-ice models \cite{C-17}.

In the case of the six-vertex model, within the various possible
choices of fixed boundary conditions, domain wall boundary conditions
\cite{K-82} are the most studied. The arctic curve is known
essentially only for the case of a square domain
\cite{CP-08,CP-09,CPZj-10}, although some recent progress has been
made about the derivation of the arctic curve for the model in domains
of more generic shape, from the knowledge of some suitable boundary
correlation function, and of its asymptotic behaviour in the scaling
limit \cite{CS-16}.

In the present paper we consider the six-vertex model in an L-shaped
domain. The partition function of the model has been evaluated in
\cite{CP-07b}.  At the free-fermion point, it can be expressed as the
partition function of some discrete log-gas \cite{J-00,P-13}. The
corresponding free energy has been studied in \cite{CP-13,CP-15}.

The analytic determination of the arctic curve of the model in an
L-shaped domain, even only at the free-fermion point, is a nontrivial
problem. In the free-fermion case, one could in principle use the
general approach developed for dimer models in
\cite{KO-06,KOS-06,KO-07}.  These papers establish the existence of
the arctic curve, and its analyticity, under general hypotheses which
apply also here. They also provide a method for the analytic determination of
the curve. However, while their method is quite amenable in
the case of a triangular lattice (lozenge tilings), it becomes rather
cumbersome for the square lattice (domino tilings). 

Here, instead, we 
resort to the recently developed `Tangent method' \cite{CS-16}.  For
this we need to calculate a suitable boundary correlation function of
the six-vertex model in an L-shaped domain, and to estimate its
asymptotic behaviour in the scaling limit. While most of the
derivation can be carried out for generic choices of the Boltzmann
weights, at the moment we are able to perform the aforementioned 
asymptotic evaluation only at the free-fermion point, the crucial limitation 
being the absence of an equivalent of
the following Proposition \ref{Prop1} in the generic case.

The paper is organized as follows. We start by defining the model and
introduce the correlation functions of interest for our purposes.
Next we recall the Tangent method, which allows to determine 
the arctic curve from the knowledge of the asymptotic
behaviour of some specific boundary correlation function.  This is
evaluated in terms of the so-called
generalized emptiness formation probability (GEFP), introduced in \cite{CPS-16},
and expressed here, at the free-fermion point, in terms of the one-point
correlation function of a discrete log-gas 
model. The asymptotic behaviour of the boundary correlation function
is then evaluated, by resorting to standard techniques of random
matrix models. Finally, using the Tangent method, the arctic curve of
the model is obtained. We also show that the 
result is in excellent agreement with the numerical evaluation of the limit shape
at finite size.

\section{The six-vertex model on an L-shaped domain}\label{sec.themodel}

In this section we recall basic facts about the six-vertex model with
domain wall boundary conditions on an L-shaped domain. We define
the GEFP, a rather general correlation function of the six-vertex
model on the usual $N\times N$ lattice with domain wall boundary
conditions, which can be specialized to describe the partition
function and some useful correlation functions in the case of the L-shaped
domain. We formulate the problem of determination of the arctic curve
arising in the scaling limit and set up some notation for parametrising the
geometry of the domain.

\subsection{The lattice, configurations and weights}

\begin{figure}[b]
\centering
\begin{tikzpicture}[scale=.5]
\draw [thick] (0.2,1)--(1.8,1);
\draw [thick] (1,0.2)--(1,1.8);
\draw [thick] [->] (0.5,1)--(.6,1);
\draw [thick] [->] (1.5,1)--(1.6,1);
\draw [thick] [->] (1,0.5)--(1,.6);
\draw [thick] [->] (1,1.5)--(1,1.6);
\node at (1,-.5) {$w_1$};
\end{tikzpicture}
\quad
\begin{tikzpicture}[scale=.5]
\draw [thick] (0.2,1)--(1.8,1);
\draw [thick] (1,0.2)--(1,1.8);
\draw [thick] [->] (0.5,1)--(.4,1);
\draw [thick] [->] (1.5,1)--(1.4,1);
\draw [thick] [->] (1,0.5)--(1,.4);
\draw [thick] [->] (1,1.5)--(1,1.4);
\node at (1,-.5) {$w_2$};
\end{tikzpicture}
\quad
\begin{tikzpicture}[scale=.5]
\draw [thick] (0.2,1)--(1.8,1);
\draw [thick] (1,0.2)--(1,1.8);
\draw [thick] [->] (0.5,1)--(.6,1);
\draw [thick] [->] (1.5,1)--(1.6,1);
\draw [thick] [->] (1,0.5)--(1,.4);
\draw [thick] [->] (1,1.5)--(1,1.4);
\node at (1,-.5) {$w_3$};
\end{tikzpicture}
\quad
\begin{tikzpicture}[scale=.5]
\draw [thick] (0.2,1)--(1.8,1);
\draw [thick] (1,0.2)--(1,1.8);
\draw [thick] [->] (0.5,1)--(.4,1);
\draw [thick] [->] (1.5,1)--(1.4,1);
\draw [thick] [->] (1,0.5)--(1,.6);
\draw [thick] [->] (1,1.5)--(1,1.6);
\node at (1,-.5) {$w_4$};
\end{tikzpicture}
\quad
\begin{tikzpicture}[scale=.5]
\draw [thick] (0.2,1)--(1.8,1);
\draw [thick] (1,0.2)--(1,1.8);
\draw [thick] [->] (0.5,1)--(.6,1);
\draw [thick] [->] (1.5,1)--(1.4,1);
\draw [thick] [->] (1,0.5)--(1,.4);
\draw [thick] [->] (1,1.5)--(1,1.6);
\node at (1,-.5) {$w_5$};
\end{tikzpicture}
\quad
\begin{tikzpicture}[scale=.5]
\draw [thick] (0.2,1)--(1.8,1);
\draw [thick] (1,0.2)--(1,1.8);
\draw [thick] [->] (0.5,1)--(.4,1);
\draw [thick] [->] (1.5,1)--(1.6,1);
\draw [thick] [->] (1,0.5)--(1,.6);
\draw [thick] [->] (1,1.5)--(1,1.4);
\node at (1,-.5) {$w_6$};
\end{tikzpicture}
\caption{The six vertex configurations and their weights.}
\label{fig-6Vertices}
\end{figure}
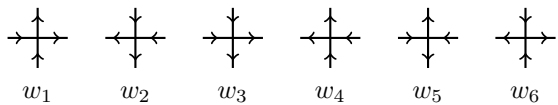

The states of the six-vertex model are configurations of arrows pointing
along the edges of a square lattice, and satisfying the condition that
at each vertex the numbers of incoming and outgoing arrows are equal.
This condition, known as \textit{ice rule}, selects six possible
vertex configurations. These are listed in Fig.~\ref{fig-6Vertices},
together with the corresponding Boltzmann weights, $w_i$,
$i=1,\dots,6$.

\begin{figure}

\usetikzlibrary{decorations.pathreplacing}
\begin{tikzpicture}[scale=.5]
\draw [thick] (0.2,1)--(7.8,1);
\draw [thick] (0.2,2)--(7.8,2);
\draw [thick] (0.2,3)--(7.8,3);
\draw [thick] (0.2,4)--(7.8,4);
\draw [thick] (0.2,5)--(7.8,5);
\draw [thick] (3.2,6)--(7.8,6);
\draw [thick] (3.2,7)--(7.8,7);
\draw [thick] (1,0.2)--(1,5.8);
\draw [thick] (2,0.2)--(2,5.8);
\draw [thick] (3,0.2)--(3,5.8);
\draw [thick] (4,0.2)--(4,7.8);
\draw [thick] (5,0.2)--(5,7.8);
\draw [thick] (6,0.2)--(6,7.8);
\draw [thick] (7,0.2)--(7,7.8);
\draw [thick] [->] (.5,1)--(.4,1);
\draw [thick] [->] (.5,2)--(.4,2);
\draw [thick] [->] (.5,3)--(.4,3);
\draw [thick] [->] (.5,4)--(.4,4);
\draw [thick] [->] (.5,5)--(.4,5);
\draw [thick] [->] (3.5,6)--(3.4,6);
\draw [thick] [->] (3.5,7)--(3.4,7);
\draw [thick] [->] (1,5.5)--(1,5.4);
\draw [thick] [->] (2,5.5)--(2,5.4);
\draw [thick] [->] (3,5.5)--(3,5.4);
\draw [thick] [->] (4,7.5)--(4,7.4);
\draw [thick] [->] (5,7.5)--(5,7.4);
\draw [thick] [->] (6,7.5)--(6,7.4);
\draw [thick] [->] (7,7.5)--(7,7.4);
\draw [thick] [->] (7.5,1)--(7.6,1);
\draw [thick] [->] (7.5,2)--(7.6,2);
\draw [thick] [->] (7.5,3)--(7.6,3);
\draw [thick] [->] (7.5,4)--(7.6,4);
\draw [thick] [->] (7.5,5)--(7.6,5);
\draw [thick] [->] (7.5,6)--(7.6,6);
\draw [thick] [->] (7.5,7)--(7.6,7);
\draw [thick] [->] (1,.5)--(1,.6);
\draw [thick] [->] (2,.5)--(2,.6);
\draw [thick] [->] (3,.5)--(3,.6);
\draw [thick] [->] (4,.5)--(4,.6);
\draw [thick] [->] (5,.5)--(5,.6);
\draw [thick] [->] (6,.5)--(6,.6);
\draw [thick] [->] (7,.5)--(7,.6);
\draw [decorate,decoration={brace}]
(3.9,8.1) -- (7.1,8.1) node [midway,yshift=9pt] {$r$};
\draw [decorate,decoration={brace}]
(2.9,5.9) -- (2.9,7.1) node [midway,xshift=-9pt] {$s$};
\draw [decorate,decoration={brace,mirror}]
(8.1,0.9) -- (8.1,7.1) node [midway,xshift=9pt] {$N$};
\draw [decorate,decoration={brace,mirror}]
(.9,-.1) -- (7.1,-.1) node [midway,yshift=-10pt] {$N$};
\node at (4,-2) {(a)};
\end{tikzpicture}
\qquad
\begin{tikzpicture}[scale=.5]
\draw [thick] (0.2,1)--(7.8,1);
\draw [thick] (0.2,2)--(7.8,2);
\draw [thick] (0.2,3)--(7.8,3);
\draw [thick] (0.2,4)--(7.8,4);
\draw [thick] (0.2,5)--(7.8,5);
\draw [thick] (0.2,6)--(7.8,6);
\draw [thick] (0.2,7)--(7.8,7);
\draw [thick] (1,0.2)--(1,7.8);
\draw [thick] (2,0.2)--(2,7.8);
\draw [thick] (3,0.2)--(3,7.8);
\draw [thick] (4,0.2)--(4,7.8);
\draw [thick] (5,0.2)--(5,7.8);
\draw [thick] (6,0.2)--(6,7.8);
\draw [thick] (7,0.2)--(7,7.8);
\draw [thick] [->] (.5,1)--(.4,1);
\draw [thick] [->] (.5,2)--(.4,2);
\draw [thick] [->] (.5,3)--(.4,3);
\draw [thick] [->] (.5,4)--(.4,4);
\draw [thick] [->] (.5,5)--(.4,5);
\draw [thick] [->] (.5,6)--(.4,6);
\draw [thick] [->] (.5,7)--(.4,7);
\draw [thick] [->] (1,7.5)--(1,7.4);
\draw [thick] [->] (2,7.5)--(2,7.4);
\draw [thick] [->] (3,7.5)--(3,7.4);
\draw [thick] [->] (4,7.5)--(4,7.4);
\draw [thick] [->] (5,7.5)--(5,7.4);
\draw [thick] [->] (6,7.5)--(6,7.4);
\draw [thick] [->] (7,7.5)--(7,7.4);
\draw [thick] [->] (7.5,1)--(7.6,1);
\draw [thick] [->] (7.5,2)--(7.6,2);
\draw [thick] [->] (7.5,3)--(7.6,3);
\draw [thick] [->] (7.5,4)--(7.6,4);
\draw [thick] [->] (7.5,5)--(7.6,5);
\draw [thick] [->] (7.5,6)--(7.6,6);
\draw [thick] [->] (7.5,7)--(7.6,7);
\draw [thick] [->] (1,.5)--(1,.6);
\draw [thick] [->] (2,.5)--(2,.6);
\draw [thick] [->] (3,.5)--(3,.6);
\draw [thick] [->] (4,.5)--(4,.6);
\draw [thick] [->] (5,.5)--(5,.6);
\draw [thick] [->] (6,.5)--(6,.6);
\draw [thick] [->] (7,.5)--(7,.6);
\draw [thick] [->] (1.5,7)--(1.4,7);
\draw [thick] [->] (1.5,6)--(1.4,6);
\draw [thick] [->] (2.5,7)--(2.4,7);
\draw [thick] [->] (2.5,6)--(2.4,6);
\draw [thick] [->] (3.5,7)--(3.4,7);
\draw [thick] [->] (3.5,6)--(3.4,6);
\draw [thick] [->] (1,6.5)--(1,6.4);
\draw [thick] [->] (2,6.5)--(2,6.4);
\draw [thick] [->] (3,6.5)--(3,6.4);
\draw [thick] [->] (1,5.5)--(1,5.4);
\draw [thick] [->] (2,5.5)--(2,5.4);
\draw [thick] [->] (3,5.5)--(3,5.4);
\draw [decorate,decoration={brace}]
(3.9,8.1) -- (7.1,8.1) node [midway,yshift=9pt] {$r$};
\draw [decorate,decoration={brace}]
(-.1,5.9) -- (-.1,7.1) node [midway,xshift=-9pt] {$s$};
\draw [decorate,decoration={brace,mirror}]
(8.1,0.9) -- (8.1,7.1) node [midway,xshift=9pt] {$N$};
\draw [decorate,decoration={brace,mirror}]
(.9,-.1) -- (7.1,-.1) node [midway,yshift=-10pt] {$N$};
\node at (4,-2) {(b)};
\end{tikzpicture}
\caption{The L-shaped domain with domain wall boundary conditions:
(a) The domain with a cut-off corner, (b) Equivalent arrow
configuration on the original lattice.  Here $N=7$, $r=4$, and
$s=2$.}
\label{fig-2Domains}
\end{figure}
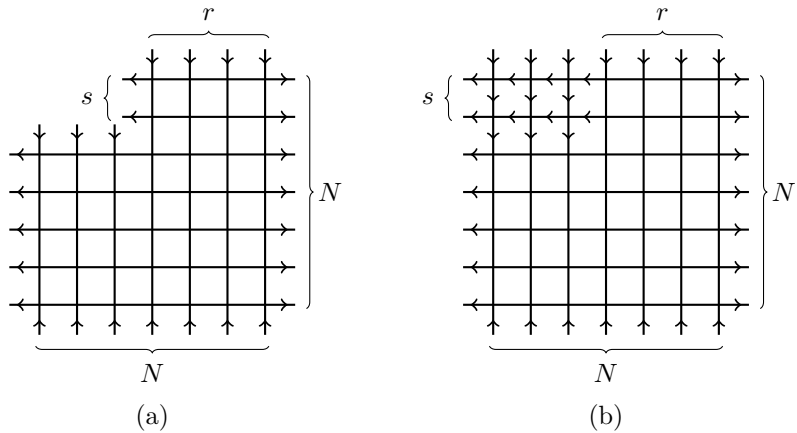

The L-shaped domain can be defined as a square domain with a
rectangular portion removed from one of the corners, see
Fig.~\ref{fig-2Domains}a.  Specifically, the square domain is the
finite square lattice obtained from the intersection of $N$ horizontal
and $N$ vertical lines (we will just say `the $N \times N$ lattice'
for `the $N\times N$ square domain of the square lattice'). The
L-shaped domain is obtained by removing a
rectangular portion of the lattice, of size $s\times (N-r)$,
from the top-left corner of the square. The interesting range is 
$r+s\leq N$ since otherwise there are no admissible arrow 
configurations.

For future convenience, we label the vertices of the square domain as
follows: we associate the lattice coordinates $(j,k)$ to the vertex at
the intersection between the $j$th vertical line, counting from the
right, and the $k$th horizontal line, counting from top.
Correspondingly, the vertices of the L-shaped domain have labels
$(j,k)$, with $j=1,\dots,r$ for $k=1,\dots,s$, and $j=1,\dots,N$ for
$k=s+1,\dots,N$.

We impose  domain wall boundary conditions by fixing all horizontal
(respectively, vertical) arrows on external edges as outgoing
(incoming). For $s=0$, one has the usual $N\times N $ lattice with
domain wall boundary conditions, introduced in \cite{K-82}.  We denote
the partition function of the six-vertex model in the $L$-shaped
domain with domain wall boundary conditions by $Z_{N,r,s}$. When
$s$ or $N-r$ is zero, we use the standard notation $Z_N$.

Under the choice of domain wall boundary conditions, due to the ice
rule, there is an obvious correspondence between the six-vertex model
on the L-shaped domain and the model on the $N\times N$ square
lattice, conditioned to have all vertex configurations of type 2 in
the top-left $s\times (N-r)$ rectangle, see Fig.~\ref{fig-2Domains}b.

It is well known that (see, e.g., \cite{BL-13}), in presence of domain
wall boundary conditions, one can restrict with no loss of generality
to Boltzmann weights that are invariant under reversal of arrows, that is 
$w_1=w_2$, $w_3=w_4$, and $w_5=w_6$. In this paper
we consider only the case in which they also obey the free-fermion condition
$w_1w_2+w_3w_4=w_5w_6$. In this case, and up to a global rescaling, 
one is left with a single real 
parameter, and we adopt the following parameterization of the Boltzmann weights:
\begin{equation}\label{param}
w_1=w_2=\sqrt{1-\alpha},
\qquad w_3=w_4=\sqrt{\alpha},
\qquad w_5=w_6=1,\qquad \alpha\in[0,1].
\end{equation}

The construction above can be translated into the language of dimer
models, using the well-known correspondence between the six-vertex
model with domain wall boundary conditions and the domino tilings of
the Aztec diamond \cite{EKLP-92}. The case of the six-vertex model on
an $N\times N$ square lattice with domain wall boundary conditions
corresponds to the Aztec diamond of order $N$; analogously, the model
on the L-shaped domain corresponds to the Aztec diamond with a cut-off
corner \cite{CP-13}.

\subsection{The GEFP and boundary correlation function}

In \cite{CPS-16}, a rather general and flexible nonlocal correlation
function, called \textit{generalized emptiness formation probability}
(GEFP), was introduced.  For the six-vertex model on a square domain,
it describes the probability of having an $s$-tuple of horizontal
edges (one edge per line, in the first $s$ horizontal lines, with
corresponding column indices forming a weakly ordered sequence) all in
a given state.

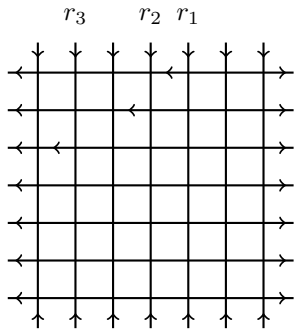
\begin{figure}

\begin{tikzpicture}[scale=.5]
\draw [thick] (0.2,1)--(7.8,1);
\draw [thick] (0.2,2)--(7.8,2);
\draw [thick] (0.2,3)--(7.8,3);
\draw [thick] (0.2,4)--(7.8,4);
\draw [thick] (0.2,5)--(7.8,5);
\draw [thick] (0.2,6)--(7.8,6);
\draw [thick] (0.2,7)--(7.8,7);
\draw [thick] (1,0.2)--(1,7.8);
\draw [thick] (2,0.2)--(2,7.8);
\draw [thick] (3,0.2)--(3,7.8);
\draw [thick] (4,0.2)--(4,7.8);
\draw [thick] (5,0.2)--(5,7.8);
\draw [thick] (6,0.2)--(6,7.8);
\draw [thick] (7,0.2)--(7,7.8);
\draw [thick] [->] (.5,1)--(.4,1);
\draw [thick] [->] (.5,2)--(.4,2);
\draw [thick] [->] (.5,3)--(.4,3);
\draw [thick] [->] (.5,4)--(.4,4);
\draw [thick] [->] (.5,5)--(.4,5);
\draw [thick] [->] (.5,6)--(.4,6);
\draw [thick] [->] (.5,7)--(.4,7);
\draw [thick] [->] (1,7.5)--(1,7.4);
\draw [thick] [->] (2,7.5)--(2,7.4);
\draw [thick] [->] (3,7.5)--(3,7.4);
\draw [thick] [->] (4,7.5)--(4,7.4);
\draw [thick] [->] (5,7.5)--(5,7.4);
\draw [thick] [->] (6,7.5)--(6,7.4);
\draw [thick] [->] (7,7.5)--(7,7.4);
\draw [thick] [->] (7.5,1)--(7.6,1);
\draw [thick] [->] (7.5,2)--(7.6,2);
\draw [thick] [->] (7.5,3)--(7.6,3);
\draw [thick] [->] (7.5,4)--(7.6,4);
\draw [thick] [->] (7.5,5)--(7.6,5);
\draw [thick] [->] (7.5,6)--(7.6,6);
\draw [thick] [->] (7.5,7)--(7.6,7);
\draw [thick] [->] (1,.5)--(1,.6);
\draw [thick] [->] (2,.5)--(2,.6);
\draw [thick] [->] (3,.5)--(3,.6);
\draw [thick] [->] (4,.5)--(4,.6);
\draw [thick] [->] (5,.5)--(5,.6);
\draw [thick] [->] (6,.5)--(6,.6);
\draw [thick] [->] (7,.5)--(7,.6);
\draw [thick] [->] (4.5,7)--(4.4,7);
\draw [thick] [->] (3.5,6)--(3.4,6);
\draw [thick] [->] (1.5,5)--(1.4,5);
\node at (5,8.5) {$r_1$};
\node at (4,8.5) {$r_2$};
\node at (2,8.5) {$r_3$};
%
\end{tikzpicture}
\caption{The configuration of arrows on the $N\times N$ lattice,
whose probability is described by the GEFP. Here $N=7$, $s=3$, and
$(r_1,r_2,r_3)=(3,4,6)$.}
\label{fig-gefp}
\end{figure}

More precisely, enumerating the horizontal lines of the $N\times N$
lattice from the top and the vertical lines from the right, we choose
$s$ edges, $e_1,\dots,e_s$, $1\leq s \leq N$, with edge $e_k$,
$k=1,\dots,s$, located on the $k$th horizontal line, and between the
$r_k$th and $(r_{k}+1)$th vertical lines. We require the $r_k$'s to
form a weakly increasing sequence $1\leq r_1\leq \dots\leq r_s\leq N$.
We denote by $G_{N,s}^{(r_1,\dots,r_s)}$ the probability of observing
all arrows on the horizontal edges $e_1,\dots,e_s$ to be pointing
left, see Fig.~\ref{fig-gefp}a.  Clearly, this is also the
probability of having vertex configurations of type 2 at all sites
$(j,k)$ with $r_k< j\leq N$, $k=1,\dots,s$. 

In \cite{CPS-16} a multiple integral representation was derived for the
GEFP (see equations (5.4) and (5.6) in that paper).  Under the
free-fermion condition and with parameterization
\eqref{param}, this representation reads
\begin{equation}\label{MIRGEFPFF}
G_{N,s}^{(r_1,\dots,r_s)}=(-1)^s
\oint_{C_0}\cdots\oint_{C_0}
\prod_{j=1}^{s}\frac{(\alpha z_j+1-\alpha)^{N-j}}{z_j^{r_j}(z_j-1)^{s-j+1}}
\prod_{1\leq j<k \leq s}^{} (z_j-z_k)
\,\frac{\rmd^s z}{(2\pi \rmi)^s}.
\end{equation}
In the special case $r_1=\dots=r_s=r$ the GEFP reduces to the usual
emptiness formation probability (EFP) of the six-vertex model with
domain wall boundary conditions, introduced in \cite{CP-08}. 

We emphasize that, although the GEFP has been defined for the model on a
square domain, by suitably specializing the values of the $r_j$'s, it
actually provides closed form expressions for the partition function
and some correlation functions of the model on certain classes of more
general domains. For example, the EFP essentially describes the quotient of the partition
functions of the model on the L-shaped domain and the original
$N\times N$ lattice, namely, the following relations holds:
\begin{equation}\label{ZNrs}
Z_{N,r,s}= \frac{Z_N}{w_2^{s(N-r)}}G_N^{(r,\ldots,r)}=
\frac{1}{(1-\alpha)^{s(N-r)/2}}G_N^{(r,\ldots,r)}.
\end{equation}
Here the first equality is valid for arbitrary weights while the
second is specific for the case of free-fermion weights, with
parameterization \eqref{param}, when the partition functions evaluates
simply to $Z_N=1$. Thus \eqref{MIRGEFPFF}, due to \eqref{ZNrs}, also
provides a multiple integral representation for the partition
function.

\begin{figure}

\begin{tikzpicture}[scale=.5]
\draw [thick] (0.2,1)--(7.8,1);
\draw [thick] (0.2,2)--(7.8,2);
\draw [thick] (0.2,3)--(7.8,3);
\draw [thick] (0.2,4)--(7.8,4);
\draw [thick] (0.2,5)--(7.8,5);
\draw [thick] (3.2,6)--(7.8,6);
\draw [thick] (3.2,7)--(7.8,7);
\draw [thick] (1,0.2)--(1,5.8);
\draw [thick] (2,0.2)--(2,5.8);
\draw [thick] (3,0.2)--(3,5.8);
\draw [thick] (4,0.2)--(4,7.8);
\draw [thick] (5,0.2)--(5,7.8);
\draw [thick] (6,0.2)--(6,7.8);
\draw [thick] (7,0.2)--(7,7.8);
\draw [thick] [->] (.5,1)--(.4,1);
\draw [thick] [->] (.5,2)--(.4,2);
\draw [thick] [->] (.5,3)--(.4,3);
\draw [thick] [->] (.5,4)--(.4,4);
\draw [thick] [->] (.5,5)--(.4,5);
\draw [thick] [->] (3.5,6)--(3.4,6);
\draw [thick] [->] (3.5,7)--(3.4,7);
\draw [thick] [->] (1,5.5)--(1,5.4);
\draw [thick] [->] (2,5.5)--(2,5.4);
\draw [thick] [->] (3,5.5)--(3,5.4);
\draw [thick] [->] (4,7.5)--(4,7.4);
\draw [thick] [->] (5,7.5)--(5,7.4);
\draw [thick] [->] (6,7.5)--(6,7.4);
\draw [thick] [->] (7,7.5)--(7,7.4);
\draw [thick] [->] (7.5,1)--(7.6,1);
\draw [thick] [->] (7.5,2)--(7.6,2);
\draw [thick] [->] (7.5,3)--(7.6,3);
\draw [thick] [->] (7.5,4)--(7.6,4);
\draw [thick] [->] (7.5,5)--(7.6,5);
\draw [thick] [->] (7.5,6)--(7.6,6);
\draw [thick] [->] (7.5,7)--(7.6,7);
\draw [thick] [->] (1,.5)--(1,.6);
\draw [thick] [->] (2,.5)--(2,.6);
\draw [thick] [->] (3,.5)--(3,.6);
\draw [thick] [->] (4,.5)--(4,.6);
\draw [thick] [->] (5,.5)--(5,.6);
\draw [thick] [->] (6,.5)--(6,.6);
\draw [thick] [->] (7,.5)--(7,.6);
\draw [thick] [->] (5,6.5)--(5,6.6);
\node at (5,8.5) {$l$};
%
\end{tikzpicture}
\caption{Boundary correlation function $H_{N,r,s}^{(l)}$ in the case 
$N=7$, $r=4$, $s=2$, and $l=3$.}
\label{fig-HNrs}
\end{figure}
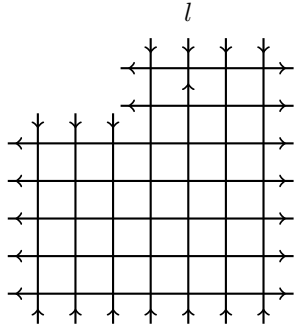

Below we shall be interested in a particular boundary correlation
function for the six-vertex model in an L-shaped domain.  To define
this function let us consider the first row of vertical edges, between
the first two horizontal lines. We note that, as a consequence of the
ice-rule and domain wall boundary conditions, the corresponding arrows
are all pointing down, except one. We denote by $H^{(l)}_{N,r,s}$ the
probability of observing this sole up arrow exactly on the $l$th
vertical edge, $l=1,\dots,r$, see Fig.~\ref{fig-HNrs}.  We introduce
the corresponding generating function,
\begin{equation}\label{generating}
h_{N,r,s}(w)= \sum_{l=1}^{r}
H_{N,r,s}^{(l)}w^{l-1},\qquad 
h_{N,r,s}(1)=1.
\end{equation}
This function plays a crucial role in what follows.

It is clear that the boundary correlation function $H^{(l)}_{N,r,s}$
is closely connected to the GEFP, upon suitable specialization of the
$r_j$'s. Indeed, for $l\leq r$, the quotient $G_{N,s}^{(l,r,\dots,r)}/G_{N,s}^{(r,r,\dots,r)}$
gives the probability of
observing, for the model on the L-shaped domain, a left-pointing arrow
on the first horizontal edge in the $l$th column. The boundary correlation function
$H^{(l)}_{N,r,s}$ is nothing but the lattice derivative in $l$ of this
probability:
\begin{equation}\label{HNrsl}
H_{N,r,s}^{(l)}=\frac{1}{G_{N,s}^{(r,\dots,r)}}
\left[G_{N,s}^{(l,r,\dots,r)}-G_{N,s}^{(l-1,r,\dots,r)}\right],
\qquad l=1,\dots,r.
\end{equation}
The generating function $h_{N,r,s}(w)$ can similarly be expressed in terms
of the GEFP, with some simplifications occurring at the level of the 
integrand in \eqref{MIRGEFPFF}.

An analogous quantity could be defined for the refinement position on
a different boundary side: in the $(s+1)$th row of vertical edges,
restricted to the last $N-r$ columns, there is at most one arrow
pointing up (and in fact, in the limit of large $N$, almost surely
one). In order to investigate the associated statistics, one may
study the lattice derivative (in $l$) of the correlation function
$G_{N,s+1}^{(r,\dots,r,l)}/G_{N,s}^{(r,\dots,r)}$, with 
$r \leq l \leq N$. As in this paper we are considering only the
free-fermionic case, in which the exact knowledge of a portion
of the curve implies the knowledge of all the curve, by analytic
continuation, we can avoid the study of this second combination
(although, in fact, at the level of the study of the log-gas, this
would require only minor modifications).

\subsection{Scaling limit, Arctic ellipse, and two regimes}

The phase separation phenomena in the six-vertex model
take place in the scaling limit, which is  
performed by sending $N\to\infty$, and
simultaneously rescaling the lattice coordinates $(j,k)$ such that: $j/N=x$, 
$k/N=y$, with $(x,y)\in[0,1]^2$ now being continuous coordinates. In the case of
the L-shaped domain, the parameters $r$ and $s$ are rescaled as well,
and we set
\begin{equation}
R=\frac{r}{s},\qquad Q=\frac{N-r-s}{s},
\end{equation}
where $R \geq 1$ (in order for the statistical ensemble to be
nontrivial) and, without loss of generality, $Q\geq 0$ (the case $Q<0$
may be obtained by symmetry).  We use $s$, rather than $N$, as the
main scaling parameter since the former naturally appears in the
discrete log-gas description of the model.

\begin{figure}

\begin{tikzpicture}[scale=.4]
\draw [thick] (1,1)--(10,1);
\draw [thick] (10,1)--(10,10);
\draw [thick] (1,1)--(1,7);
\draw [thick] (1,7)--(5,7);
\draw [thick] (5,7)--(5,10);
\draw [thick] (5,10)--(10,10);
\draw [<->] (-.5,10)--(10,10)--(10,-.5);
\node at (10.6,10.5) {$0$};
\node at (10.6,-.5) {$y$};
\node at (-.5,10.6) {$x$};
\draw [very thin] (1,10)--(1,10.2);
\draw [very thin] (10,1)--(10.2,1);
\node at (1,10.6) {1};
\node at (10.6,1) {1};
\draw [very thin] (5,10)--(5,10.2);
\draw [very thin] (10,7)--(10.2,7);
\node at (5,10.6) {$\xi_x$};
\node at (10.8,7) {$\xi_y$};
\end{tikzpicture}
\caption{The portion of the $\mathbb{R}^2$ plane corresponding to the
L-shaped domain in the scaling limit.}
\label{fig-ScalingLimit}
\end{figure}
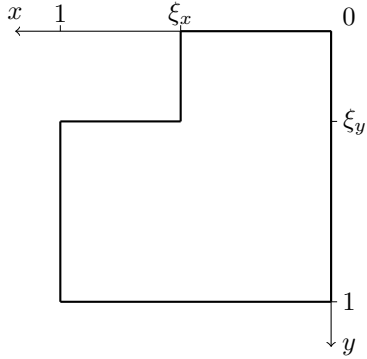

The $R$ and $Q$ fully describe the geometry of
L-shaped region in the scaling limit. An alternate useful 
parametrization is given by 
the coordinates of the bottom-right vertex of the rectangular cut-off
corner of the L-shaped domain:
\begin{equation}\label{xixxiy}
\xi_x=\frac{R}{R+Q+1}, \qquad 
\xi_y=\frac{1}{R+Q+1}.
\end{equation}
Thus, in the scaling limit, the L-shaped domain is rescaled into the
region $\{[0,\xi_x]\times[0,\xi_y]\}\cup \{[0,1]\times[\xi_y,1]\}$ of
the $\mathbb{R}^2$ plane, see Fig.~\ref{fig-ScalingLimit}. Note that
the coordinates $x$ and $y$ have origin in correspondence of the
top-right vertex of the L-shaped domain, and are oriented leftward and
downward, respectively. This unconventional choice is done in order to
match with the labeling of the rows and columns in the discrete
lattice.

It is known \cite{CP-13,CP-15} that the model undergoes a third-order
phase transition as the cut-off rectangle is large enough to touch the
Arctic ellipse of the model on the original (unmodified) lattice. More
precisely, the phase transition occurs, for given values of $Q$ and
$\alpha$, at $R=\Rc(Q,\alpha)$:
\begin{equation}\label{Rc}
\Rc=\frac{\left(1+\sqrt{\alpha(1+Q)}\right)^2}{1-\alpha}.
\end{equation} 
This curve splits the space of parameters 
$(R,Q,\alpha)\in[1,\infty)\times [0,\infty)\times[0,1]$
into two regions, which we
call Regime I, when $R\in [\Rc,\infty)$, and Regime II, $R\in [1,\Rc]$.

In terms of the coordinates $\xi_x$ and $\xi_y$, the value $\Rc$
corresponds to one arc of the ellipse
\begin{equation}\label{AE}
\frac{(1-x-y)^2}{1-\alpha}+\frac{(x-y)^2}{\alpha}=1,
\end{equation}
tangent to the four sides of the unit square, and specifically to that
arc which connects tangency points $(x,y)=(\alpha,0)$ and
$(x,y)=(1,1-\alpha)$.  This arc is also described by the equation
$\sqrt{y}=\sqrt{(1-\alpha)x}-\sqrt{\alpha(1-x)}$.

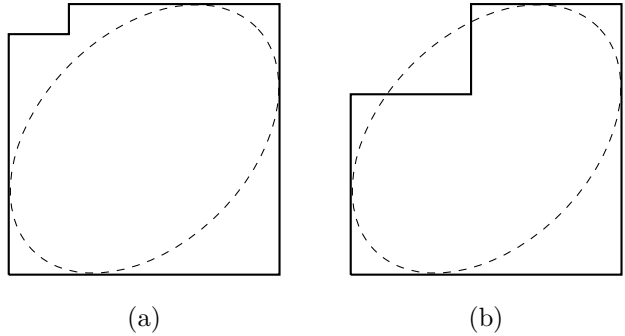
\begin{figure}

\begin{tikzpicture}[scale=.4]
\draw [thick] (1,1)--(10,1)--(10,10)--(3,10)--(3,9)--(1,9)--(1,1);
\draw [dashed,rotate=45] (7.8,0) ellipse (5.25 and 3.5);
\node at (5.5,-.5) {(a)};
\end{tikzpicture}
\quad
\begin{tikzpicture}[scale=.4]
\draw [thick] (1,1)--(10,1)--(10,10)--(5,10)--(5,7)--(1,7)--(1,1);
\draw [dashed,rotate=45] (7.8,0) ellipse (5.25 and 3.5);
\node at (5.5,-.5) {(b)};
\end{tikzpicture}
\caption{The two regimes: (a) Regime I, (b) Regime II.}
\label{fig-2Regimes}
\end{figure}

In other words, \eqref{AE}
is nothing but the Arctic ellipse of the original six-vertex model (at
the free-fermion point). Then, Regime I corresponds to the situation where
the cut-off rectangle lies totally outside the Arctic ellipse; in
Regime II the cut-off rectangle `penetrates' the interior of
the Arctic ellipse, see Fig.~\ref{fig-2Regimes}. 
In the former case the arctic curve is not modified by the restricted geometry,
while in latter case  it must be deformed into 
some new curve, whose determination is the goal here. 

\begin{figure}
\includegraphics[scale=.75]{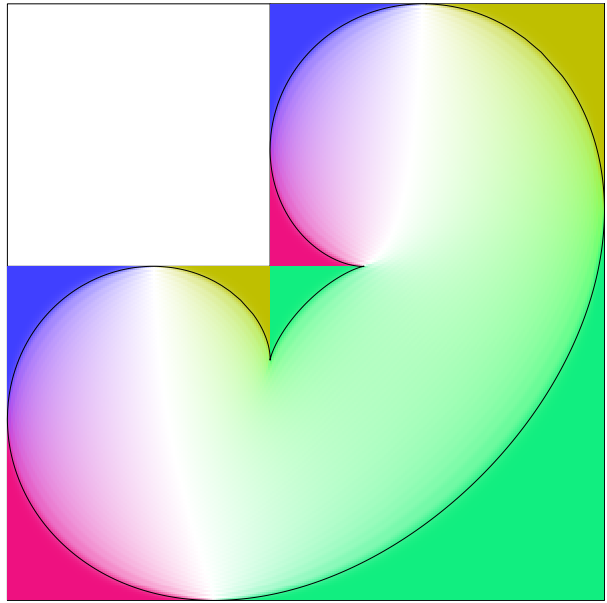}
\caption{Illustration of the edge-inclusion probabilities, in the
domain with $N=300$ and $s=N-r=132$, done according to the `Limit
shape' procedure described in Appendix \ref{appendixnum}. In black,
the plot of the arctic curve \eqref{xypm} with the same ratios $r/N$
and $s/N$.}
\label{fig.numplot2}
\end{figure}

In the case of the L-shaped domain with domain wall boundary
conditions, phase separation phenomena and the emergence of a
nontrivial limit shape should be expected, on the basis of the general
results of \cite{KO-06,KOS-06,KO-07}. On top of this, the
phenomenon is clearly observed in numerics, see Fig.~\ref{fig.numplot2}.
The numerics presented in this picture has been generated
basing on the generalized domino shuffling algorithm \cite{P-03}, 
see Appendix \ref{appendixnum} for details.

The analytic determination of the arctic curve of the model is a
nontrivial problem. In the free-fermion case, one could in principle
resort to the general approach developed for dimer models in
\cite{KO-06,KOS-06,KO-07}. However, for the square lattice, the
characteristic polynomial of the corresponding Kasteleyn matrix is of
degree higher than one, and makes it difficult to work out explicit
results beyond the case of the square domain.

Here we shall resort to the Tangent method, that determines the Arctic
curve of the six-vertex model in a generic domain as the geometric
caustic of a family of straight lines, that are completely determined
by the behaviour of the boundary correlation function in the scaling
limit \cite{CS-16}. We emphasize that this paper provides just one
application of the general method, which instead applies, in
principle, to a wide class of domains and to generic Boltzmann
weights.

\section{The Tangent method}

In the section we illustrate how the Tangent method may be applied to
the six-vertex model on an L-shaped domain with domain wall boundary
conditions. Our main object of study is the one-point boundary
correlation function, whose asymptotic behaviour in the scaling limit
determines the arctic curve. The boundary correlation function may be
represented in terms of a discrete log-gas. Correspondingly, we show
that its scaling limit behaviour is described by (the functional
inverse of) the resolvent associated to the discrete log-gas in the
thermodynamic limit.

\subsection{Parametric equation for the arctic curve}\label{sec_3.1}

In the case of the L-shaped domain, the Tangent method gives the
following recipe: the arc of the arctic curve subtended by the corner
with lattice coordinates $(r,1)$, i.e., $(\xi_x,0)$
in the scaling limit, can be expressed in parametric form $x=x(w)$,
$y=y(w)$, with $w\in[1,\infty)$, as the solution of the linear system
of equations
\begin{equation}\label{rspe}
F(w;x,y)=0, \qquad\partial_w F(w;x,y)=0,
\end{equation}
with 
\begin{equation}\label{rspe2}
F(w;x,y) =x - M(w) y - \Phi(w),\qquad
M(w)\equiv\frac{w}{(w-1)(\alpha w +1-\alpha)}.
\end{equation}
Here the function $\Phi(w)=\Phi(w;R,Q,\alpha)$ is defined as follows:
\begin{equation}\label{defPhi}
\Phi(w)
=\lim_{N,r,s\to\infty} \frac{1}{N}w\partial_w \log h_{N,r,s}(w),
\end{equation}
where $h_{N,r,s}(w)$ is
the generating function of the boundary correlation function $H_{N,r,s}^{(l)}$,
see \eqref{generating}.

Note that the recipe may be equivalently formulated as follows: the
above considered arc of arctic curve is the geometric caustic of the
one-parameter family \eqref{rspe2} of straight lines in the
$(x,y)$-plane, in the parameter $w\in[1,\infty)$. Also, the values
$w=1$, and $w\to\infty$ correspond to the two points of contact of
the considered arc with the two sides of the L-shaped domain:
$\{(x,0),\ x\in[0,\xi_x]\}$ and $\{(\xi_x,y),\ y\in[0,\xi_y]\}$,
respectively.

The above recipe follows from the `path description' of the six-vertex
model, and can be applied to all those models whose configurations can
be rephrased in terms of (directed) non-intersecting lattice paths
(although the paths are allowed to `osculate', that is, they may have
a contact-point interaction). The path description highlights some
conservation law of the models and displays their underlying fermionic
(but not necessarily free-fermionic) character.

These paths can also be interpreted as level lines of a certain height
function, whose value at the boundary is fixed. This height function,
in the scaling limit, may converge to some limit shape.  Frozen
regions are associated to portions of the limit shape which are flat.
In this picture, and under the mild assumption that the typical
distance between consecutive paths is $o(N)$, it is clear that the
limiting behaviour of the outmost path describes (some portion of) the
arctic curve. The idea is then to condition one end-point of the most
external lattice path to some distant lattice point. In the scaling
limit, the path is expected to follow only a portion of the outer
shell of path (that is, the arctic curve), then to escape it
tangentially, and, free from the influence of other paths (as the
paths only interact locally), to continue toward the prescribed
end-point along a straight line.

This heuristic picture has an analytic counterpart at the level of the
boundary correlation function associated to prescribing the position
of the end-point, and modulo the very reasonable and strongly
supported, but still unproven, tangency assumption, may be developed
rigorously. A standard saddle-point analysis leads to the Tangent
method's recipe.  According to the above interpretation, for each
value of $w\in[1,\infty)$, in the scaling limit, the slope of the
  straight portion of the out-most path is $1/M(w)$, while the
  quantity $\Phi(w)$ is essentially the value of concentration in $l$
  for the boundary correlation function $H_{N,r,s}^{(l)}$; that is, in
  the generating function $h_{N,r,s}(w)$ (see (\ref{generating})), the
  most relevant summands are concentrated around $l=\Phi(w)$, within a
  window of the order $\sqrt{N}$.  For a full description of the
  Tangent method, with many examples, see \cite{CS-16}. For further
  applications, see \cite{DfL-18,DfG-18}.

In principle, the above procedure should be performed for each corner
of the domain, to determine the corresponding subtended arc of the
arctic curve. However, in the considered case of free-fermion
Boltzmann weights, the arctic curve is known to be algebraic
\cite{KO-06}, allowing for the possibility of extending one arc to a
full component of the arctic curve, just by analytic continuation.
In our case this accounts to extending the range of parameter $w$ 
to the whole real axis, and taking both determinations 
in a certain square root expression.

Clearly, the implementation of the Tangent method requires the
explicit knowledge of $\Phi(w)$, and thus the calculation of the
boundary correlation function $H_{N,r,s}^{(l)}$, and the evaluation of
the asymptotic behaviour of the corresponding generating function in
the scaling limit. This is our main task below.

\subsection{Discrete log-gas representation}

Different representations, in terms of multiple integrals, or of
determinants, can be worked out for $H_{N,r,s}^{(l)}$, and for the
corresponding generating function, $h_{N,r,s}(w)$. The most convenient
one for our purposes is in terms of a discrete log-gas:

\begin{proposition}\label{Prop1}
For the generating function $h_{N,r,s}(w)$
the following representation is valid:
\begin{equation}\label{HNrs}
h_{N,r,s}(w)=\frac{w^{r-1}I_{N,r,s}(u)}{I_{N,r,s}(1)}.
\end{equation}
Here, the variables $u$ and $w$ are related by
\begin{equation}\label{ualpha}
u=\frac{\alpha w +1-\alpha}{w},
\end{equation}
and the function $I_{N,r,s}(u)$ is given as
\begin{multline}\label{INrsu}
I_{N,r,s}(u)=
\sum_{m_1,\dots,m_s=0}^{r-1}
\prod_{j=1}^s\mu_{N-r-s}^\alpha(m_j)
\prod_{1\leq j<k\leq s} (m_k-m_j)^2
\\ \times
\oint_{C_{m_1,\ldots,m_s}}\frac{\sigma(u,z)}{\prod_{j=1}^s(z-m_j)}
\frac{\rmd z}{2\pi\rmi}, 
\end{multline}
where
\begin{equation}\label{measure}
\mu_q^\alpha(m)=\alpha^{m}\binom{q+m}{q},\qquad m\in\mathbb{N}_0,
\end{equation}
and
\begin{equation}\label{sigma}
\sigma(u,z)=(s-1)!\frac{u^{r+s-z-2}}{(1-u)^{s-1}}.
\end{equation}
The integration in \eqref{INrsu} is over a simple counterclockwise
oriented contour enclosing all $m_j$'s, and no other singularity of
the integrand.
\end{proposition}

The proof of proposition \ref{Prop1} is provided in appendix
\ref{appendixprop}, and it goes along the lines of what is done in
\cite{P-13} in absence of refinement, that is, for the quantity $I_{N,r,s}(1)$.

Concerning the statements of Proposition \ref{Prop1} several remarks
are in order.  First, we note that $w=(1-\alpha)/(u-\alpha)$, and so
$w\to 1$ as $u\to 1$, hence \eqref{HNrs} reproduces the normalization
condition $h_{N,r,s}(1)=1$, see \eqref{generating}.

Second, even though the function $\sigma(u,z)$ is singular as $u\to
1$, the integral in \eqref{INrsu} is regular at $u=1$. Indeed, for any
fixed $s$, the integral over $z$ in \eqref{INrsu} can be estimated as
$u\to 1$ as follows:
\begin{multline}
\oint_{C_{m_1,\ldots,m_s}} \frac{\sigma(u,z)}{\prod_{j=1}^s (z-m_j)}
\frac{\rmd z}{2\pi\rmi}
= (s-1)! \frac{u^{s+r-2}}{(1-u)^{s-1}} \oint_{C_{m_1,\ldots,m_s}}
\frac{\rme^{-z\log u}}{\prod_{j=1}^s (z-m_j)} \frac{\rmd z}{2\pi\rmi}
\\ 
\sim \frac{u^{s+r-2}}{(1-u)^{s-1}} (-\log u)^{s-1} \sim 1.
\end{multline}

Third, the quantity $I_{N,r,s}(1)$ reads
\begin{equation}\label{INrs}
I_{N,r,s}(1)=
\sum_{m_1,\dots,m_s=0}^{r-1}
\prod_{j=1}^s\mu_{N-r-s}^\alpha(m_j)
\prod_{1\leq j<k\leq s}
(m_k-m_j)^2.
\end{equation}
From this expression it follows that $I_{N,r,s}\equiv I_{N,r,s}(1)$
can be viewed as the partition function of a discrete log-gas confined
within a finite interval, note the condition $0\leq m_j<r$,
$j=1,\dots,s$, for the particle coordinates.  Correspondingly, the
quantity $I_{N,r,s}(u)$ can be viewed as a particular correlation
function of the discrete log-gas defined by \eqref{INrs}.  The
discrete weight \eqref{measure} is that of the Meixner
polynomials. The partition function and free energy of this log-gas
with discrete measure \eqref{measure} have been studied in details in
\cite{J-00}, see also \cite{BKMM-07}. The role and consequences of the
condition $0\leq m_j<r$, $j=1,\dots,s$ on the behaviour of the free
energy have been discussed in \cite{CP-13,CP-15}.

\subsection{Relation between $\Phi(w)$ and the resolvent of the log-gas}

To proceed further, we need to evaluate the quantity
$\Phi(w)$ defined in \eqref{defPhi}.
Recalling \eqref{HNrs}, we may write
\begin{equation}\label{Phiprem}
\Phi(w)
=\frac{1}{R+Q+1}\left(
R+w \frac{\partial u}{\partial w} \lim_{N,r,s\to\infty}\frac{1}{s}
\partial_u\log I_{N,r,s}(u)\right),
\end{equation}
and thus we need to estimate the large
$s$ behaviour of the correlation function $I_{N,r,s}(u)$ for the
discrete log-gas with measure \eqref{measure} in a scaling limit 
with $r/s$ and $N/s$ fixed.  

Let us focus first on the partition function $I_{N,r,s}$ of the same
discrete log-gas. Its large $s$ behaviour may be determined in the
saddle-point approximation.  The standard procedure is to rescale the
eigenvalues by a factor $s$, namely $m_j\to s z_j$.  After rescaling, the
sums in \eqref{INrs} can be reinterpreted as Riemann sums, and, in the
large-$s$ limit, replaced by integrals:
\begin{equation}\label{INrs2}
I_{N,r,s}\propto
\int_{0}^{R}\dots\int_{0}^{R}\rmd z_1\cdots \rmd z_s
\prod_{j=1}^s\mu^{\alpha}_{N-r-s}(\lfloor sz_j\rfloor)
\prod_{1\leq j<k \leq s}(z_k-z_j)^2
\end{equation}

Now the usual saddle-point analysis for Random Matrix models can be
applied, provided that one imposes a suitable additional constraint
accounting for the discreteness of the $m_j$'s \cite{DK-93}, see
\cite{CP-13,CP-15} for full details on the specific example of
\eqref{INrs}. The solution $\{\tilde{z}_j\}_{j=1,\dots,s}$ of the set of
saddle-point equations associated to the multiple integrals in
\eqref{INrs2} is encoded in the resolvent $W(z)$, defined as
\begin{equation}\label{spe}
W(z)=\lim_{s \to\infty}\frac{1}{s}\sum_{j=1}^s\frac{1}{z-\tilde{z}_j}.
\end{equation}
In the case of $I_{N,s,r}(u)$, rescaling the log-gas `coordinates'
$m_j\to sz_j$ and simultaneously replacing $z\to sz$, we get
\begin{multline}\label{INrs3}
I_{N,r,s}(u)\propto
\int_{0}^{R}\dots\int_{0}^{R}\rmd z_1\cdots \rmd z_s
\prod_{j=1}^s\mu^{\alpha}_{N-r-s}(\lfloor sz_j\rfloor)
\prod_{1\leq j<k \leq s}(z_k-z_j)^2 
\\ \times
\oint_{C_{z_1,\ldots,z_s}}
\frac{\sigma(u,sz)}{\prod_{j=1}^s(z-z_j)}\rmd z,
\end{multline}
The crucial point is that the set of saddle-point equations for the
$z_j$'s remains the same as for the case of $I_{N,r,s}$, and the
corresponding solution is still encoded in $W(z)$. However, there is an
additional saddle-point equation, relative to the variable $z$, which,
recalling \eqref{sigma}, reads
\begin{equation}\label{spez}
-\log u =W(\tilde{z}).
\end{equation}
In other words, the saddle-point value for the extra integration variable is
given by the functional inverse of the resolvent:
\begin{equation}\label{spez2}
\tilde{z} =W^{-1}(-\log u).
\end{equation}
The inversion relation \eqref{spez2} appears in several settings, see,
e.g., \cite{ZJ-98}.

Differentiating the logarithm of \eqref{INrs3} with respect to $u$,
we find 
\begin{equation}\label{dulogI}
\partial_u \log I_{N,r,s}(u)=
\langle \partial_u\log\sigma(u,sz) \rangle
\end{equation}
where brackets $\langle \cdot \rangle$ denote the expectation value with
respect to the measure associated to $I_{N,r,s}(u)$, see
\eqref{INrs3}.  Taking into account that
\begin{equation}
\partial_u\log\sigma(u,sz) =\frac{s}{1-u}+ \frac{s+r-sz}{u} 
\end{equation}
and using that in the scaling limit $\langle z \rangle = \tilde{z}$,
we get
\begin{equation}
\lim_{N,r,s\to\infty}\frac{1}{s}\partial_u \log I_{N,r,s}(u)=
\frac{1}{1-u}+\frac{1+R-W^{-1}(-\log u)}{u},
\end{equation}
where we also have made use of \eqref{spez2}. Finally, 
using \eqref{Phiprem} and \eqref{HNrs}, we get
\begin{equation}\label{phiQ}
\Phi(w)
=\frac{1}{(R+Q+1)u}
\left[R\alpha +\frac{u-\alpha}{u-1}
+(u-\alpha) W^{-1}(-\log u)\right],
\end{equation}
where variables $u$ and $w$ are related by \eqref{ualpha}.

Thus, the evaluation of $\Phi(w)$ has boiled down to that of (the
functional inverse of) $W(z)$, that is the resolvent associated to the
discrete log-gas \eqref{INrs}. The expressions of this resolvent for
the various regimes has been worked out in \cite{CP-13,CP-15}.

\section{Equation for the arctic curve}

In this section we focus on details of derivation of the arctic curve
using the results of Ref.~\cite{CP-15} on the explicit form of the
resolvent $W(z)$. We consider various cases, in order of increasing
complexity: we start with Regime I, next we consider Regime II for a
symmetric domain (the cut-off rectangle is a square) that corresponds to
$Q=0$, and, finally, we treat the case of Regime II in full generality
($Q\geq 0$).

\subsection{Regime I}

In this case, $ R>\Rc$, and $ R$ does not
enter the expression of the resolvent, which reads (see \cite{CP-13}):
\begin{multline}\label{WI}
W(z)=-\log\sqrt\alpha
-\log\frac{\sqrt{a(z-b)}+\sqrt{b(z-a)}}{\sqrt{(b-a)z}}
\\
\mp
\log\frac{\sqrt{(a+Q)(z-b)}+\sqrt{(b+Q)(z-a)}}{\sqrt{(b-a)z}},
\end{multline}
where
\begin{equation}
a=\frac{\left(1-\sqrt{\alpha(1+Q)}\right)^2}{1-\alpha},\qquad
b=\frac{\left(1+\sqrt{\alpha(1+Q)}\right)^2}{1-\alpha}.
\end{equation}
In \eqref{WI} the choice of the sign depends on the value of the
parameter $Q$, with the critical value $\Qc=\alpha^{-1}-1$
corresponding to the case where $a=0$; the minus sign corresponds to
$Q<\Qc$ and the plus sign to $Q>\Qc$. Solving equation $-\log u=W(z)$
for $z$, we get the following solution valid in both cases:
\begin{equation}\label{z_of_u}
z=-\frac{[1-\alpha(1+Q)]u+\alpha Q}{(u-1)(u-\alpha)}, 
\qquad Q\in[0,\infty).
\end{equation}
The function \eqref{phiQ} reads
\begin{equation}\label{Phiw}
\Phi(w)=\frac{\alpha}{u}= \frac{\alpha w}{\alpha w+1-\alpha}.
\end{equation}
Note that dependence from $Q$ and $R$ cancel out in the expression of
$\Phi(w)$.  Plugging the result into \eqref{rspe2} and solving of the
linear system \eqref{rspe} yields
\begin{equation}\label{arctic}
x=\frac{\alpha w^2}{\alpha w^2+1-\alpha},
\qquad y=\frac{\alpha(1-\alpha)(w-1)^2}{\alpha w^2+1-\alpha}.
\end{equation}

Here, according to the recipe prescribed in Section \ref{sec_3.1} the
parameter $w$ should run over the interval $[1,\infty)$.  Actually,
the rectangular portion removed from the top-left corner to build
the L-shaped domain constitutes a forbidden region for the family of
lines \eqref{rspe2}. As a result, the parameter $w$ is, by 
construction, allowed to run only over the interval $[1,w_0)$, where
$w_0$ is the largest of the two solutions of:
\begin{equation}\label{w0}
(w-1)(\alpha w+1-\alpha)\xi_x-w\xi_y-\alpha w (w-1)=0.
\end{equation}
Condition \eqref{w0} selects, within the family of lines
\eqref{rspe2}, with $\Phi(w)$ given by \eqref{Phiw}, the two lines
lines passing through the point of coordinates $(\xi_x,\xi_y)$. 

Apart from this technical detail, the already mentioned fact that in
the presently considered free-fermion case the arctic curve is
guaranteed to be algebraic \cite{KO-06} allows anyway to extend the
range of $w$ to the whole real axis, $w\in\mathbb{R}$, and
correspondingly to describe the whole arctic curve.  Indeed,
eliminating $w$ in \eqref{arctic} yields the Arctic ellipse
\eqref{AE}, as expected for the Regime I.

The implicit form of equation of the arctic curve, given by \eqref{AE}
can also be directly recovered by considering the polynomial
$P(u)=u(u-1)F(w)$, where $F(w)=F(w;x,y)$ is the function defined in
\eqref{rspe2}.  Explicitly, $P(u)$ reads
\begin{equation}
P(u)=u(u-1)x+(u-\alpha)y-\alpha(u-1).
\end{equation}
Since system \eqref{rspe} implements the condition that two zeroes of
the function $F(w)$ should coincide, we can directly impose this
condition by requiring that the discriminant of $P(u)$ vanish.  This
immediately gives \eqref{AE}.

\subsection{Regime II, symmetric domain}

In Regime II the resolvent $W(z)$ has in
general a rather complicate expression, and it is convenient to focus
first on the case of a symmetric domain, where 
the cut-off rectangle is actually a square. In this case $s=N-r$, 
that is $Q=0$. The Regime II means that
the other geometric parameter $R$ is in the range 
\begin{equation}\label{RcQ0}
R: 1\leq R< \Rc,\qquad
\Rc=\frac{1+\sqrt{\alpha}}{1-\sqrt{\alpha}}.
\end{equation}
The resolvent reads (see \cite{CP-13}):
\begin{equation}\label{WII}
W(z)=-\log\sqrt\alpha
+\log\frac{z}{z-R}
-2\log\frac{\sqrt{a(z-b)}+\sqrt{b(z-a)}}
{\sqrt{(R-a)(z-b)}+\sqrt{(R-b)(z-a)}},
\end{equation}
where
\begin{equation}\label{abII}
a=\frac{\big(\sqrt{ R+1}
-\sqrt{(R-1)\sqrt{\alpha}}\big)^2}{2(1+\sqrt{\alpha})},
\qquad
b=\frac{\big(\sqrt{ R+1}
+\sqrt{(R-1)\sqrt{\alpha}}\big)^2}{2(1+\sqrt{\alpha})}.
\end{equation}

The relation $-\log u=W(z)$ can be written in the form:
\begin{equation}\label{WIIbis}
u=\sqrt{\alpha}
\frac{\big(\sqrt{(R-a)(z-b)}-\sqrt{(R-b)(z-a)}\big)
\big(\sqrt{b(z-a)}+\sqrt{a(z-b)}\big)}
{\big(\sqrt{(R-a)(z-b)}+\sqrt{(R-b)(z-a)}\big)
\big(\sqrt{b(z-a)}-\sqrt{a(z-b)}\big)}.
\end{equation}
Using \eqref{abII} and taking into account \eqref{RcQ0}, 
one can simplify it to 
\begin{equation}
u=\sqrt{\alpha}
\frac{(1+\sqrt{\alpha})\big(\sqrt{(z-a)(z-b)}-\sqrt{ab}\big)+(1-\sqrt{\alpha})z}
{(1+\sqrt{\alpha})\big(\sqrt{(z-a)(z-b)}+\sqrt{ab}\big)-(1-\sqrt{\alpha})z}
\end{equation} 
Solving for $z$, we get two solutions:
\begin{multline}\label{zpm}
z=\frac{R}{2}-\frac{(1-\alpha)u}{2(u-\alpha)(u-1)}
\pm\frac{(u-\sqrt{\alpha})\sqrt{R^2(u-\alpha)(u-1)+(1+\sqrt{\alpha})^2u}}
{2(u-\alpha)(u-1)}.
\end{multline}  
Apparently, these solutions represent two branches of the same
function $W^{-1}(-\log u)$, which determines the arctic curve; a
particular choice of the sign in \eqref{zpm} corresponds to a portion
of the arctic curve, via the function (see \eqref{phiQ})
\begin{equation}
\Phi(w)=\frac{1}{(R+1)u}\left[R\alpha+\frac{u-\alpha}{u-1}+(u-\alpha) z\right],
\end{equation}
which reads
\begin{equation}\label{PhiII}
\Phi(w)=\frac{\alpha}{2u}+\frac{\xi_x}{2}+\frac{\xi_y(u-\alpha)}{2u(u-1)}
\pm\frac{(u-\sqrt{\alpha})
\sqrt{\xi_x^2(u-\alpha)(u-1)+\xi_y^2(1+\sqrt{\alpha})^2u}}{2u(u-1)}.
\end{equation}
Here, we have employed the notations for the coordinates of the
bottom-right vertex of the cut-off rectangle, $\xi_x=R/(R+1)$,
$\xi_y=1/(R+1)$, see \eqref{xixxiy}, and we also recall that the
variables $u$ and $w$ are related by \eqref{ualpha}.

\begin{figure}
\centering
\includegraphics[scale=.45, angle=180]{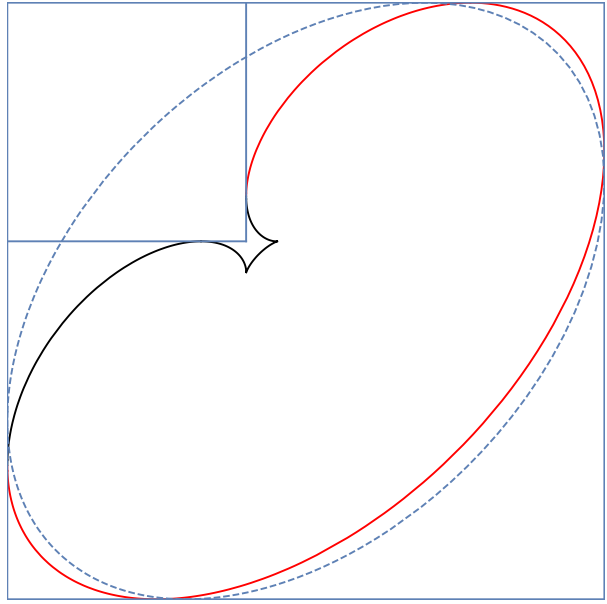}
\caption{Plot of the two components $\mathcal{C}_-$ and
$\mathcal{C}_+$, in red (grey in b/w printing) and black,
respectively, as given in \eqref{xypm}; here
$\alpha=0.3$ (corresponding to $\Rc\simeq 3.42$), and
$ R=1.5$. The dashed line shows, for comparison, the Arctic
ellipse \eqref{AE}.}
\label{fig-plot1}
\end{figure}

\begin{figure}
\includegraphics[scale=.75]{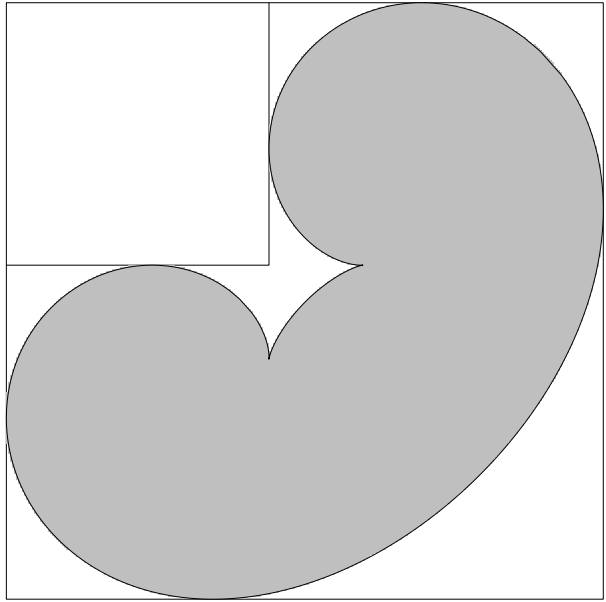}
\caption{In grey, numerical determination of the arctic curve, in the
domain with $N=300$ and $s=N-r=132$, done according to the `Arctic
Curve' procedure described in Appendix \ref{appendixnum}. In black,
the plot of the arctic curve \eqref{xypm} with the same ratios $r/N$
and $s/N$.}
\label{fig.numplot1}
\end{figure}

To investigate the resulting arctic curve, let us denote by 
$\Phi_{+}(w)$ and $\Phi_{-}(w)$ the function in \eqref{PhiII}
taken with the plus and minus signs, respectively. Consider 
two different parametric families of straight lines:
\begin{equation}\label{Fpm}
F_\pm(w)=x-M(w)y-\Phi_\pm(w).
\end{equation} 
Solving the corresponding system 
of equations \eqref{rspe} in $x$ and $y$, we obtain 
\begin{equation}\label{xypm}
\begin{split}
x_{\pm}(w)&=-\frac{M(w)\Phi_\pm^\prime(w)}{M^\prime(w)}+\Phi_{\pm}(w),
\\
y_{\pm}(w)&=-\frac{\Phi_{\pm}^\prime(w)}{M^\prime(w)},
\end{split}
\end{equation}
where the prime denotes differentiation with respect to $w$.  This is
the parametric form of the two branches $\mathcal{C}_+$ and
$\mathcal{C}_-$ of the arctic curve, given respectively by
$(x_{+}(w),y_{+}(w))$ and $(x_{-}(w),y_{-}(w))$, with
$w\in\mathbb{R}$. These expressions make it possible to produce plots
of the arctic curve; Fig.~\ref{fig-plot1} shows an example for
particular values of the parameters. In Fig.~\ref{fig.numplot1},
another example of the curve, corresponding to a different
choice of parameters is plotted against numerics. See 
Appendix \ref{appendixnum} for further details.

The obtained arctic curve has six points of contact with the
boundary of the L-shaped domain, and two cusps, for a total of eight
special points.  Starting from the bottom side of the boundary, and proceeding 
counterclockwise, the first four points correspond to the values
$w=-(1-\alpha)/\alpha$, $0$, $1$, $\infty$, in $(x_{+}(w),y_{+}(w))$,
while the next four correspond to the same values of $w$, in
$(x_{-}(w),y_{-}(w))$. It is apparent from inspection of formula
\eqref{rspe2} that these values of $w$ indeed correspond to 
(alternately in cyclic order) vanishing
or diverging values for the slope of the arctic curve.

The equation of the arctic curve can also be obtained in an implicit
form. Because of the geometry of the curve, it is convenient to
introduce \emph{diagonal} coordinates
\begin{equation}\label{lightcone}
z_1=x-y, \qquad 
z_2=1-x-y. 
\end{equation}
Introduce the functions (we use here the variable $u$ in the argument):
\begin{equation}
\wt F_\pm(u)\equiv 2 u(u-1) F_\pm(w(u)).
\end{equation}
Explicitly, these functions read
\begin{multline}\label{Ftilde}
\wt F_\pm(u) = (z_1-z_2-\xi_x)u(u-1)-(z_1+z_2+\xi_y)(u-\alpha)+(u-\alpha)u
\\
\pm(u-\sqrt{\alpha})\sqrt{\xi_x^2(u-\alpha)(u-1)+\xi_y^2(1+\sqrt{\alpha})^2u}.
\end{multline}
Consider the following polynomial of degree 4 in $u$:
\begin{equation}\label{Pu}
P(u)= \wt F_{+}(u) \wt F_{-}(u).
\end{equation}
Clearly, the equation for the arctic curve can be derived by 
requiring that this polynomial has vanishing 
discriminant. More precisely, denoting by $D(P)$
the discriminant of $P(u)$, it is fairly easy to see 
from the formulas above, that $D(P)$ is a polynomial 
of degree 8 in $z_1$, $z_2$. Moreover, one can easily verify
(using a symbolic manipulation software) that 
it has the following structure: 
\begin{equation}\label{factorized}
D(P)=\left(z_1-\frac{1+\sqrt{\alpha}}{2}+\xi_y\right)^2
\mathcal{A}(z_1,z_2),
\end{equation}
where $\mathcal{A}(z_1,z_2)$ is a polynomial of degree 6 in $z_1$ and
$z_2$. The equation $\mathcal{A}(z_1,z_2)=0$ is the desired implicit
equation of the arctic curve.

The straight line of equation described by the first factor in
\eqref{factorized} is precisely that sole line which is tangent to the
arctic curve at two distinct points. Indeed, let $w_0\in [1,\infty)$
be the (unique) solution to the condition $M(w_0)=1$, we have
\begin{equation}
w_0=\frac{2\Rc}{\Rc-1},\qquad 
\Phi_\pm(w_0)=\frac{R\Rc-1}{( R+1)(\Rc+1)}=\frac{1+\sqrt{\alpha}}{2}-\xi_y.
\end{equation}
This component arises as a side effect of the Tangent method.

The explicit expression for $\mathcal{A}(z_1,z_2)$ which is of degree 6
in $z_1$ and $z_2$ is rather lengthy and reported in appendix
\ref{appendixresult} (some of its 28 coefficients appear to be zero, but
most of them are complicate polynomials in $R$, $\alpha$).  At $R=1$,
corresponding to $\xi_x=\xi_y=1/2$, that is to a cut-off square of side
$1/2$, the equation $\mathcal{A}(z_1,z_2)=0$ factorizes into two Arctic ellipses
and two coinciding straight lines, tangent to both of them, as expected.

\subsection{Regime II, generic domain}

We now turn to the case $Q>0$, where the cut-off portion is a
rectangle rather than a square. In this case, in Regime II, the
resolvent reads
\begin{multline}
W(z)=-\log\sqrt\alpha
+2\frac{\log{\sqrt{(R-a)(z-b)}+\sqrt{(R-b)(z-a)}}}
{\sqrt{(b-a)(z-R)}}
\\
-\log\frac{\sqrt{(a+Q)(z-b)}+\sqrt{(a+Q)(z-a)}}{\sqrt{(b-a)z}}
\\
\mp\log\frac{\sqrt{a(z-b)}+\sqrt{a(z-a)}}{\sqrt{(b-a)z}},
\end{multline}
where the parameters $a$ and $b$, $0\leq a<b\leq R$, are to be found
from the equations\footnote{In \cite{CP-15} there is a misprint in
(2.20), (2.22) and (3.8): In the first factor of the first equation
the replacement $a\leftrightarrow b$ should be made.}:
\begin{equation}\label{ab-eqns-RII}
\begin{split}
\sqrt{\alpha}\frac{\sqrt{R-a}-\sqrt{R-b}}{\sqrt{R-a}+\sqrt{R-b}}
\frac{\sqrt{b}\pm\sqrt{a}}{\sqrt{b+Q}-\sqrt{a+Q}}
&=1,\\
\frac{\pm\sqrt{ab}+\sqrt{(a+Q)(b+Q)}-Q}{2}+\sqrt{(R-a)(R-b)}
&=1.
\end{split}
\end{equation}
Here, the $\pm$ signs correspond to what is called in
Ref.~\cite{CP-15} the Regime IIA, $Q\leq \Qc$, and Regime IIB,
$Q>\Qc$, respectively, where the value $Q=\Qc$ is determined by the
condition $a=0$. In contrast to the $Q=0$ case, explicit expressions
for $a$ and $b$ at $Q\ne 0$ are cumbersome functions of $R$, $Q$ and
$\alpha$, and in the most compact form they can be written as
\begin{equation}\label{aAAbAA}
a=A_{+}+A_{-}-2\sqrt{A_{+}A_{-}},\qquad
b=A_{+}+A_{-}+2\sqrt{A_{+}A_{-}}.
\end{equation}
where 
\begin{equation}
\begin{split}
A_{+}&=(R+Q+1) \frac{(1+\eta)(1+R\eta)\left[1+(R+Q)\eta\right]}{\left[2+Q+(2R+Q)\eta\right]^2},
\\
A_{-}&=(R-1)
\frac{(1-\eta)(1+Q+R\eta)
  \left[1+Q+(R+Q)\eta\right]}{\left[2+Q+(2R+Q)\eta\right]^2}.
\end{split}
\end{equation}
and the parameter $\eta\in [0,1]$ is a suitable root of the (quartic) equation 
\begin{equation}\label{eq-eta}
\alpha \frac{(1+\eta)^2(1+Q+R\eta)\left[1+(R+Q)\eta\right]}
{(1-\eta)^2(1+R\eta)\left[1+Q+(R+Q)\eta\right]}=1.
\end{equation}
For the values of the parameters $R$, $Q$ and $\alpha$ belonging to the 
Regime II (i.e., for $R< \Rc$, where $\Rc=\Rc(Q,\alpha)$ 
is given by \eqref{Rc}) such 
a root always exists and it is unique \cite{CP-15}.  

The equation $-\log u =W(z)$ reads
\begin{multline}
u=\sqrt{\alpha}
\frac{\sqrt{(R-a)(z-b)}-\sqrt{(R-b)(z-a)}}
{\sqrt{(R-a)(z-b)}+\sqrt{(R-b)(z-a)}}
\\ \times 
\frac{\sqrt{(b+Q)(z-a)}+\sqrt{(a+Q)(z-b)}}
{\sqrt{b(z-a)}\mp\sqrt{a(z-b)}},
\end{multline}
that is 
\begin{equation}\label{uKL}
\frac{u}{\sqrt{\alpha}}=
\frac{K_2\sqrt{(z-a)(z-b)}+K_1 z+ K_0}
{L_2\sqrt{(z-a)(z-b)}+L_1 z+ L_0}, 
\end{equation} 
where 
\begin{equation}\label{3K3L}
\begin{split}
K_2&=\sqrt{(R-a)(b+Q)}-\sqrt{(R-b)(a+Q)},
\\
K_1&=\sqrt{(R-a)(a+Q)}-\sqrt{(R-b)(b+Q)},
\\
K_0&=a\sqrt{(R-b)(b+Q)}-b\sqrt{(R-a)(a+Q)},
\\
L_2&=\sqrt{(R-a)b}\mp\sqrt{(R-b)a},
\\
L_1&=\sqrt{(R-b)b}\mp\sqrt{(R-a)a},
\\
L_0&=\pm b\sqrt{(R-a)a}-a\sqrt{(R-b)b}.
\end{split}
\end{equation}
Solving \eqref{uKL} for $z$, we get
\begin{equation}\label{zgen}
z=\frac{M_0M_1+\left(\frac{a+b}{2}\right)M_2^2
\pm 
M_2\sqrt{(a M_1 +M_0)(b M_1+M_0)+
  \left(\frac{b-a}{2}\right)^2M_2^2}}{M_2^2-M_1^2},
\end{equation}
where $M_i=M_i(u)$ are linear functions of $u$:
\begin{equation}\label{Mi}
M_i=L_i \frac{u}{\sqrt{\alpha}}-K_i,\qquad i=0,1,2.
\end{equation}
Using the first equation in \eqref{ab-eqns-RII}, it can be shown that
\begin{equation}
M_2^2-M_1^2=\frac{(b-a)^2}{\alpha}(u-\alpha)(u-1).
\end{equation}
Hence, for the function $\Phi(w)$ defined by \eqref{phiQ}, we 
obtain the following expression:
\begin{multline}\label{newPHI}
\Phi(w)=\frac{1}{(R+Q+1)u(u-1)}
\Bigg\{
(R\alpha+1)u-\alpha(R+1)+
\alpha\frac{M_0M_1+\left(\frac{a+b}{2}\right)M_2^2}{(b-a)^2}
\\
\pm
\frac{\sqrt{\alpha}M_2}{2(b-a)}
\sqrt{\frac{4\alpha(a M_1 +M_0)(b M_1+M_0)}{(b-a)^2}+\alpha M_2^2}
\Bigg\}.
\end{multline}
Here, some terms can be simplified; for example, using just
\eqref{3K3L}, one may find that
\begin{equation}
\alpha
\frac{M_0M_1+\left(\frac{a+b}{2}\right)M_2^2}{(b-a)^2}
=R u^2+c_1u +\alpha (R-Q),   
\end{equation} 
where, however, the coefficient of the linear term, $c_1$, in contrast
to other coefficients possesses a rather bulky expression even when
\eqref{aAAbAA} is invoked.  The same can be inferred about the
coefficients of the quadratic polynomial in $u$ standing under the
square root sign in \eqref{newPHI}.

Nevertheless, the expression \eqref{newPHI} describes
the arctic curve for a generic L-shaped domain. All considerations 
made above in the $Q=0$ case extends to the present case ($Q\geq 0$) 
as well, concerning both parametric and implicit form of the curve.

Namely, denote $\Phi_{+}(w)$ and $\Phi_{-}(w)$ the function in
\eqref{newPHI} taken with the plus and minus signs, respectively, and
consider two different parametric families of straight lines described
by \eqref{Fpm}. Then \eqref{xypm} provides a parametric form of two
branches of the whole arctic curve. In producing plots of the Arctic
curve, the only difference with the $Q=0$ case is that now one has first
to obtain values of the parameters $a$ and $b$ from \eqref{aAAbAA}, by
solving equation \eqref{eq-eta} for at a given set of the main
parameters $R$, $Q$, and $\alpha$, and next to plug all the values
into \eqref{3K3L}, which determine the linear functions $M_i=M_i(u)$
defined by \eqref{Mi}.

Lastly, one can also address the problem of finding an equation which
describes the arctic curve in implicit (rather then in parametric)
form.  Here, again this equation can be found from the condition of
vanishing of the discriminant of the corresponding quartic polynomial
$P(u)$, constructed from the functions $F_\pm (w)$ by \eqref{Ftilde}
and \eqref{Pu}. The discriminant $D(P)$, similarly to
\eqref{factorized}, factors into two straight lines and the arctic
curve $\mathcal{A}(z_1,z_2)$, which is of degree 6.

\section{Acknowledgements}

We are grateful to A. Abanov, S. Chhita and F. Franchini
for interesting discussions.  We are indebted to B. Wieland for
sharing with us the code for generating uniformly sampled alternating-sign 
matrices.  We thank the Simons Center for Geometry and Physics
(SCGP, Stony Brook), research program on `Statistical Mechanics and
Combinatorics' and the Galileo Galilei Institute for Theoretical
Physics (GGI, Florence), research programs on `Statistical Mechanics,
Integrability and Combinatorics' and `Entanglement in Quantum
Systems', for hospitality and support at some stage of this work. FC
is grateful to LIPN, \'equipe Calin at Universit\'e Paris 13, for
hospitality and support at some stage of this work.  AGP and AS are
grateful to INFN, Sezione di Firenze for hospitality and support at
some stage of this work. AGP acknowledges partial support from the
Russian Science Foundation, grant \#18-11-00297.

\appendix

\section{Comparison with  finite-size results}\label{appendixnum}

In this paper we have determined the arctic curve of a free-fermionic
model.  As a result of the simplifications occurring in this case,
with respect to what it would be for a generic six-vertex model
prediction, it is much easier to perform a comparison of the result
with informations obtained by alternative methods.

In particular, through the correspondence with a model of dimer
coverings on a bipartite planar graph, at finite size, a suitable
1-point function in the bulk can be calculated, either from the
inverse Kasteleyn matrix, or, more efficiently, through a method,
devised by Propp, as part of the Urban Renewal, or Generalised Domino
Shuffling, algorithm for the exact sampling of configurations (see
\cite{P-03}, Section~3).

\begin{figure}
\includegraphics[scale=1.3]{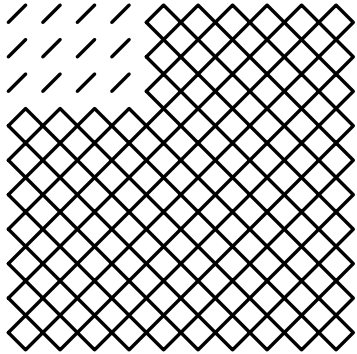}
\caption{The Aztec Diamond graph related to the L-shaped domain.}
\label{fig.graph}
\end{figure}

Our geometry is particularly adapted to the use of Propp's
algorithm. With respect to the graphical notation in \cite{P-03} (see
in particular Section~1.2), we shall just initialise the weights as in
a graph of the form shown in Fig.~\ref{fig.graph}.
Then, from the algorithm we obtain the \emph{edge-inclusion
probabilities}, that is, the probabilities $p_{ij}$, $q_{ij}$,
$r_{ij}$ and $s_{ij}$ that the edges in the plaquette of coordinates
$(i,j)$, and position NW, NE, SW and SE, respectively, are occupied in
an uniformly chosen perfect matching compatible with the domain shape
(again notations are chosen as to match with those in
\cite{P-03}). Frozen regions correspond to coordinates $(i,j)$ such
that the quadruples $(p_{ij},q_{ij},r_{ij},s_{ij})$ are equal to
$(1,0,0,0)$, $(0,1,0,0)$, etc., up to corrections exponentially small
in the size of the domain. We represent graphically these four
functions in a compact way, with two different strategies, aiming at
representing the arctic curve, or, instead, the limit shape.

In the first case, consider the combination 
\begin{equation}
x_{ij}=\frac{1}{2}(1+p_{ij}-q_{ij}-r_{ij}+s_{ij}),	
\end{equation}
associated to each plaquette, that is valued in $[0,1]$, and is near to 0 or to 1 in
the frozen regions (it is the local fraction of dimers which are
oriented diagonally, instead that anti-diagonally). We plot in gray
the plaquettes $(i,j)$ such that $x_{ij}$ is valued in
$[\eps,1-\eps]$, where $\eps=N^{-2/3}$. The scaling of this
threshold marks the change of regime between typical and atypical
local fluctuations of the arctic curve \cite{J-00}.  The choice of the
multiplicative constant 1 is of no special significance, and any other
finite constant would have produced similar results. The comparison
with our analytic prediction, shown in Figure \ref{fig.numplot1}, is
remarkably good (everywhere within one lattice spacing).

In the second case, a more refined visualization of the
edge-inclusion probabilities is obtained by associating to a
plaquette the complex number 
\begin{equation}
z_{ij}=\sqrt{p_{ij}} + \rmi\,
\sqrt{q_{ij}}- \rmi\, \sqrt{r_{ij}} - \sqrt{s_{ij}}.	
\end{equation}
This quantity is
valued in the disk of radius 1, and is exponentially near to $1$,
$\rmi$, $-1$ or $-\rmi$, if the plaquette is in a frozen region. We make a
coloured plot of the domain, with hue determined according to the
argument of $z_{ij}$, and brightness determined according to the
absolute value of $z_{ij}$ (so that the colour is near to white in
the liquid region). The data are shown in Figure \ref{fig.numplot2}.

\section{Proof of Proposition \ref{Prop1}}
\label{appendixprop}

We present here the derivation of representation \eqref{HNrs} for the
generating function $h_{N,r,s}(w)$, which is defined by
\eqref{generating} and \eqref{HNrsl}. It can be written as
\begin{equation}\label{HNrsFNrs}
h_{N,r,s}(w)= \frac{F_{N,r,s}(w)}{F_{N,r,s}(1)},
\end{equation}
where 
\begin{equation}
F_{N,r,s}(w)=\sum_{r_1=1}^{r} 
\left(G_{N,s}^{(r_1,r,\dots,r)}-G_{N,s}^{(r_1-1,r,\dots,r)}\right)
w^{r_1-1}.
\end{equation}
Note that $F_{N,r,s}(1)=G_{N,s}^{(r,\ldots,r)}$ is the EFP of the
six-vertex model with domain wall boundary conditions
($F_{N,r,s}(1)\equiv F_N^{(r,s)}$, in the notation of \cite{CP-07b}).
Change of the integration variables $z_j\mapsto x_j=(\alpha
z_j+1-\alpha)/z_j$, $j=1,\ldots,s$, in \eqref{MIRGEFPFF} yields
\begin{multline}
G_{N,s}^{(r_1,\ldots,r_s)}=(-1)^{\frac{s(s-1)}{2}}
\prod_{j=1}^{s} (1-\alpha)^{N-r_j}
\\ \times	
\oint_{C_\infty}^{} \cdots \oint_{C_\infty}^{} \prod_{j=1}^s
\frac{x_j^{N-j}}{(x_j-\alpha)^{N-r_j}(x_j-1)^{s-j+1}}
\prod_{1\leq j<k\leq s}(x_k-x_j)\,
\frac{\rmd^s x}{(2\pi\rmi)^s},
\end{multline}
where $C_\infty$ denotes a circular contour of large radius around the
origin (thus enclosing the points $x=\alpha$ and $x=1$). Hence,
\begin{multline}
F_{N,r,s}(w)=(-1)^{\frac{s(s-1)}{2}}(1-\alpha)^{(N-r)s} w^{r-1}
\\ \times	
\oint_{C_\infty}^{} \cdots \oint_{C_\infty}^{} 
\frac{x_1^{N-1}}{(x_1-\alpha)^{N-r}(x_1-1)^{s-1}(x_1-u)}
\\ \times 
\prod_{j=2}^s
\frac{x_j^{N-j}}{(x_j-\alpha)^{N-r}(x_j-1)^{s-j+1}} \prod_{1\leq j<k\leq s}(x_k-x_j)\,
\frac{\rmd^s x}{(2\pi\rmi)^s},
\end{multline}
where $u=(\alpha w +1-\alpha)/w$. Using 
\begin{equation}
\det \left[(x_{s-k+1}-\alpha)^{s-j}\right]_{j,k=1,\ldots,s}
=\prod_{1\leq j<k\leq s}^{}(x_k-x_j)
\end{equation}
we can write $F_{N,r,s}(w)$ in the form of an $s\times s$ determinant
\begin{equation}\label{FdetA}
F_{N,r,s}(w)=(-1)^{\frac{s(s-1)}{2}}(1-\alpha)^{s(s+q)} w^{r-1} \det A(u), 
\end{equation}
where the matrix $A(u)$ contains dependence on $u$ only in the last
column
\begin{equation}
A_{jk}(u)=
\begin{cases}
\displaystyle
\oint_{C_\infty} \frac{x^{r+q+k-1}}{(x-\alpha)^{q+j}(x-1)^k}\,
\frac{\rmd x}{2\pi\rmi} 
&\quad k\ne s
\\
\displaystyle
\oint_{C_\infty} \frac{x^{r+q+s-1}}{(x-\alpha)^{q+j}(x-1)^{s-1}(x-u)}\,
\frac{\rmd x}{2\pi\rmi} 
&\quad k=s,
\end{cases}
\end{equation}
and where we have set $N=r+s+q$, $q\geq 0$.

To proceed with \eqref{FdetA}, it is useful to consider first the case
$w=1$, that corresponds to $u=1$. Using
\begin{multline}\label{ident}
\oint_{C_\infty} \frac{x^{c}}{(x-\alpha)^{a}(x-\beta)^b}\,\frac{\rmd x}{2\pi\rmi} 
= \frac{1}{(a-1)!(b-1)!}\partial_\alpha^{a-1} 
\partial_\beta^{b-1} \oint_{C_\infty} \frac{x^{c}}{(x-\alpha)(x-\beta)}\,
\frac{\rmd x}{2\pi\rmi}
\\
= \sum_{m=a-1}^{c-b}\binom{m}{a-1} \binom{c-m-1}{b-1} \alpha^{m-a+1} \beta^{c-m-b}, 
\qquad a,b,c\in \mathbb{N},
\end{multline}
for the entries of the matrix $A\equiv A(1)$, upon setting $\beta=1$,
$a=q+j$, $b=k$, and $c=r+q+k-1$ and making the change $m\mapsto m+q$,
we get
\begin{equation}\label{Ajk}
A_{jk}=\sum_{m=j-1}^{r-1}\binom{m+q}{q+j-1} \binom{r+k-2-m}{k-1} \alpha^{m-j+1}.
\end{equation}
Consider now entries of a given column; since 
\begin{equation}
\binom{m+q}{q+j-1}=\frac{q!}{(q+j-1)!}\binom{m+q}{q} (m)_{j-1}, 
\end{equation}
where $(m)_{a}:=m (m-1)\dots (m-a+1)$ denotes the falling factorial,
we have
\begin{equation}
A_{jk}=\frac{q!}{(q+j-1)! \alpha^{j-1}} \wt A_{jk} 
\end{equation}
where
\begin{equation}\label{wtA}
\wt A_{jk}=\sum_{m=0}^{r-1}\binom{m+q}{q} (m)_{j-1}
\binom{r+k-2-m}{k-1} \alpha^{m},
\end{equation}
and hence
\begin{equation}
\det A= 
\frac{(q!)^s}{\prod_{j=0}^{s-1}(q+j)! } \alpha^{-\frac{s(s-1)}{2}} 
 \det \wt A.
\end{equation}
The determinant of $\wt A$ evaluates as follows 
\begin{multline}\label{detwtA}
\det \wt A 
=\sum_{m_1,\ldots,m_s=0}^{r-1}
\prod_{k=1}^{s}\binom{m_k+q}{q}\binom{r+k-2-m_k}{k-1} 
\prod_{l<k}^{}(m_k-m_l)
\alpha^{m_1+\ldots+m_s}
\\
=
\frac{(-1)^{\frac{s(s-1)}{2}}}{\prod_{j=0}^{s} j!}
\sum_{m_1,\ldots,m_s=0}^{r-1}
\prod_{k=1}^{s}\binom{m_k+q}{q} \prod_{l<k}^{}(m_k-m_l)^2
\alpha^{m_1+\ldots+m_s},
\end{multline}
where we have used the fact that $(m)_{j-1}$ is a monic polynomial of
degree $j-1$ in $m$, and, similarly, that $\binom{r+k-2-m}{k-1}$ is a
polynomial of degree $k-1$ in $m$, with the leading coefficient
$(-1)^{k-1}/(k-1)!$. In total, our calculation amounts to
\begin{multline}
F_{N,r,s}(1)
=\frac{(q!)^s}{\prod_{j=0}^{s-1}(q+j)!\prod_{j=0}^{s}j!}
\frac{(1-\alpha)^{s(s+q)}}{\alpha^{\frac{s(s-1)}{2}}}
\\ \times
\sum_{m_1,\ldots,m_s=0}^{r-1}
\prod_{j=1}^{s}\binom{m_j+q}{q}
\prod_{l<k}(m_k-m_l)^2\alpha^{m_1+\ldots+m_s}.
\end{multline}
Note that this representation may equivalently be written as
\begin{equation}\label{FNrs1}
F_{N,r,s}(1)
=\frac{(q!)^s}{\prod_{j=0}^{s-1}(q+j)!j!}
\frac{(1-\alpha)^{s(s+q)}}{\alpha^{\frac{s(s-1)}{2}}}
\det \left[\sum_{m=0}^{r-1}\binom{m+q}{q}m^{j+k-2}\alpha^m\right]_{j,k=1,\ldots,s},
\end{equation}
in agreement with \cite{J-00,P-13}.

Consider now the case of generic $w$. To apply the derivation above 
with a minimal modification, consider 
instead of the matrix $A(u)$ some matrix $B(u)$, which differs from 
$A(u)$ only in the entries of the last column, 
\begin{equation}
B_{js}(u) = \oint_{C_\infty}^{} \frac{x^{r+q}}{(x-\alpha)^{q+j}}
\left(\frac{x^{s-1}}{(x-1)^{s-1}(x-u)}+\sum_{k=1}^{s-1}\gamma_k\frac{x^{k-1}}{(x-1)^k}
\right)\, \frac{\rmd x}{2\pi \rmi},
\end{equation}
where $\gamma_k$, $k=1,\ldots,s-1$, are some constants in $x$.  Note
that $\det A(u) =\det B(u)$. For $\gamma_k=u^{s-1-k}/(u-1)^{s-k}$ the
pole at $x=1$ disappears in the integral, since
\begin{equation}
\sum_{k=1}^{s-1}\gamma_k\frac{x^{k-1}}{(x-1)^k}=
\frac{u^{s-1}}{(u-1)^{s-1}(x-u)}-\frac{x^{s-1}}{(x-1)^{s-1}(x-u)}.
\end{equation}
Therefore, with this choice of $\gamma_k$'s, and recalling \eqref{ident},
we have 
\begin{equation}
B_{js}(u)=\frac{q!}{(q+j-1)!\alpha^{j-1}}
\frac{u^{r+s-2}}{(u-1)^{s-1}}
\sum_{m=0}^{r-1}\binom{m+q}{q} (m)_{j-1} \left(\frac{\alpha}{u}\right)^m.
\end{equation}
Similarly to \eqref{wtA}, introduce matrix $\wt B(u)$,
with entries 
\begin{equation}
\wt B_{jk}(u)=
\begin{cases}
\wt A_{jk}& k\ne s\\
\displaystyle
\sum_{m=0}^{r-1} \binom{q+m}{q} (m)_{j-1}
\left(\frac{\alpha}{u}\right)^{m} & k=s.
\end{cases}
\end{equation}
We have 
\begin{equation}
\det B(u)= 
\frac{(q!)^s}{\prod_{j=0}^{s-1}(q+j)! }
\frac{u^{r+s-2}}{\alpha^{\frac{s(s-1)}{2}}(u-1)^{s-1}}
\det\wt B(u).
\end{equation}
In this case, the analogue of \eqref{detwtA} is 
\begin{multline}
\det \wt B(u) 
=\sum_{m_1,\ldots,m_s=0}^{r-1}
\prod_{k=1}^{s}\binom{m_k+q}{q}
\\ \times
\prod_{k=1}^{s-1}\binom{r+k-2-m_k}{k-1} 
\prod_{l<k}^{}(m_k-m_l)
\frac{\alpha^{m_1+\ldots+m_s}}{u^{m_s}}
\\
=
\frac{(-1)^{\frac{(s-1)(s-2)}{2}}}{\prod_{j=0}^{s-2} j!}
\sum_{m_1,\ldots,m_s=0}^{r-1}
\prod_{k=1}^{s}\binom{m_k+q}{q} \prod_{l<k}^{}(m_k-m_l)
\\ \times
\prod_{k=1}^{s-1} m_k^{k-1}
\frac{\alpha^{m_1+\ldots+m_s}}{u^{m_s}}.
\end{multline}
Symmetrizing  the summand with respect to permutations 
of $m_1,\ldots,m_s$ and substituting everything in \eqref{FdetA}, 
we get 
\begin{multline}\label{FNrsw}
F_{N,r,s}(w)
=\frac{(q!)^s}{s!\prod_{j=0}^{s-1}(q+j)!\prod_{j=0}^{s-2}j!}
\frac{(1-\alpha)^{s(s+q)}}{\alpha^{\frac{s(s-1)}{2}}} w^{r-1}
\frac{u^{r+s-2}}{(u-1)^{s-1}}
\\ \times
\sum_{m_1,\ldots,m_s=0}^{r-1}
\prod_{j=1}^{s}\binom{m_j+q}{q}
\prod_{l<k}(m_k-m_l)\alpha^{m_1+\ldots+m_s}
\\ \times
\sum_{p=1}^{s} (-1)^{p-1}
\prod_{\substack{l<k\\ l,k\ne p}}(m_k-m_l)
 u^{-m_p}.
\end{multline}
Finally, rewriting the sum over $p$ as a contour integral, we arrive
at
\begin{equation}
F_{N,r,s}(w)=\frac{(q!)^s}{\prod_{j=0}^{s-1}(q+j)!\prod_{l=0}^{s}j!}
\frac{(1-\alpha)^{s(N-r)}}{
\alpha^{s(s-1)/2}}w^{r-1}I_{N,r,s}(u)
\end{equation}
where the quantity $I_{N,r,s}(u)$ is defined in
\eqref{INrsu}. Recalling \eqref{HNrsFNrs}, the statement 
of the Proposition \ref{Prop1}, representation \eqref{HNrs},
immediately follows.

We also mention that  \eqref{FNrsw} can be written as 
\begin{equation}\label{FNrsw-det}
F_{N,r,s}(w)
=\frac{(q!)^s}{\prod_{j=0}^{s-1}(q+j)!\prod_{j=0}^{s-2}j!}
\frac{(1-\alpha)^{s(s+q)}}{\alpha^{\frac{s(s-1)}{2}}}
w^{r-1} \frac{u^{r+s-2}}{(1-u)^{s-1}} 
\det H,
\end{equation}
where the $s\times s$ matrix $H$ is
\begin{equation}
H_{jk}=
\begin{cases}
\displaystyle
\sum_{m=0}^{r-1}\binom{m+q}{q}m^{j+k-2}\alpha^m & k\ne s
\\	
\displaystyle
\sum_{m=0}^{r-1}\binom{m+q}{q}m^{j-1}	\left(\frac{\alpha}{u}\right)^m & k=s.
\end{cases}
\end{equation}
Note that, as  $w\to 1$ (that is, $u\to 1$), the expected
result \eqref{FNrs1} is reproduced from \eqref{FNrsw-det}
upon taking into account that $\det H$ has a
zero of order $(s-1)$ at $u=1$.

\section{Arctic curve for Regime II, symmetric domain}\label{appendixresult}

Here we report explicit expression for the polynomial
$\mathcal{A}(z_1,z_2)$ describing the arctic curve, for the case $Q=0$ of the
model in Regime II (symmetric L-shaped domain). The 
curve is given by the equation $\mathcal{A}(z_1,z_2)=0$ and it is of degree 6.

We first introduce properly 
scaled diagonal coordinates $Z_1$ and $Z_2$, defining them by
\begin{equation}\label{scaled-zs}
z_1=\sqrt{\alpha }Z_1,\qquad
z_2=\sqrt{1-\alpha }Z_2.
\end{equation}
Recall that the original diagonal coordinates are defined by
\eqref{lightcone}. Note, that in terms of the new coordinates the
Arctic ellipse \eqref{AE} just reads
\begin{equation}\label{AEZZ}
Z_1^2+Z_2^2=1.
\end{equation}

Next, we introduce the following parameterization for the scaling 
parameter $R\in [1,\Rc]$:
\begin{equation}
R=\frac{1+\sqrt{\alpha }\beta}{1-\sqrt{\alpha}\beta},\qquad \beta\in[0,1].  
\end{equation}
The meaning of this re-parametrization is to simplify further 
expressions for the coefficients of the arctic curve, making them 
polynomials in $\alpha$  and $\beta$.

At last, we introduce coefficients $C_{n_1n_2}$ which describe the
polynomial $A(z_1,z_2)$ appearing in \eqref{factorized}, in terms of
the coordinates \eqref{scaled-zs}
\begin{equation}
\mathcal{A}(z_1,z_2)=(1-\alpha)^2\alpha^6\sum_{0\leq n_1+n_2\leq 6} 
C_{n_1 n_2} Z_1^{n_1} Z_2^{n_2}.
\end{equation}
Note that because of the symmetry of the L-shaped domain under
reflection with respect to the North-West$/$South-East diagonal, the
arctic curve possesses the symmetry $A(z_1,-z_2)=A(z_1,z_2)$, that is,
it depends only on even powers of $z_2$, i.e., $C_{n_1n_2}=0$ if $n_2$
is odd ($n_2=1,3,5$). This excludes 12 coefficients out of 28 in
total, which describe a generic degree 6 curve.

The nonzero 16 coefficients have the following expressions:
\begingroup
\allowdisplaybreaks
\begin{align}
C_{60}&
=64 (1-\alpha)^2\left(1-2\alpha\beta+\alpha\beta^2\right)^2,
\notag\\
C_{50}&
=64 (1-\alpha )^2 \left[1-(5+2\alpha)\beta+18\alpha\beta^2
-2\alpha(4+7\alpha)\beta^3+13\alpha^2\beta^4-3\alpha^2\beta^5\right], 
\notag\\
C_{42}&
=128(1-\alpha )^2\left[1+(2-6\alpha)\beta
-2\left(1-\alpha-3\alpha^2\right)\beta^2+2(1-3\alpha)\alpha\beta^3
+\alpha^2\beta^4\right],
\notag\\
C_{40}&
=16(1-\alpha)\big[1+\alpha-\left(22-18\alpha+8\alpha^2\right)\beta
+\left(41+13\alpha-32\alpha^2+8\alpha^3\right)\beta^2
\notag\\ &\quad
-4\alpha\left(36-25\alpha-\alpha^2\right)\beta^3
+\alpha\left(52+63\alpha-85\alpha^2\right)\beta^4
-6\alpha^2(13-11\alpha)\beta^5
\notag\\ &\quad
+(15-13\alpha)\alpha^2\beta^6\big],
\notag\\
C_{32}&
=128(1-\alpha)^2\big[(1-2(2+\alpha)\beta-(4-18\alpha)\beta^2
+\left(4-6\alpha-14\alpha^2\right)\beta^3
\notag\\ &\quad
-(4-13 \alpha)\alpha\beta^4-2\alpha^2 \beta^5\big],
\notag\\
C_{30}&
=32 (1-\alpha)(1-\beta)\big[\alpha-\left(2+3\alpha+2\alpha^2\right)\beta 
+\left(22-21\alpha+19\alpha^2\right)\beta^2
\notag\\ &\quad
-\alpha\left(59-48\alpha+19\alpha^2\right)\beta^3
+\alpha\left(22+18\alpha-15\alpha^2\right)\beta^4
\notag\\ &\quad
-\alpha^2(28-17\alpha)\beta^5+\alpha^2(5-3\alpha)\beta^6\big],
\notag\\ 
C_{24}&
=64 (1-\alpha )^2 \big[1+(8-12\alpha)\beta
-2\left(4-\alpha-6\alpha^2\right)\beta^2
+4\alpha(2-3\alpha)\beta^3+\alpha^2\beta^4\big],
\notag\\ 
C_{22}&
=-32(1-\alpha)\big[1+\left(12-26\alpha+8\alpha^2\right)\beta
-\left(15-3\alpha-35\alpha^2+8\alpha^3\right)\beta^2
\notag\\ &\quad
-\left(22-96\alpha+90\alpha^2+4\alpha^3\right)\beta^3
+\left(12-25\alpha-35\alpha^2+63\alpha^3\right)\beta^4
\notag\\ &\quad
-2\alpha\left(6-26\alpha+23\alpha^2\right)\beta^5
-\alpha^2(6-7\alpha)\beta^6\big],
\notag\\ 
C_{20}&
=4(2-\alpha )\alpha -16\alpha\left(9-8\alpha+\alpha^2\right)\beta
+4\left(24+78\alpha-36\alpha^2-42\alpha^3+4\alpha ^4\right)\beta^2
\notag\\ &\quad
-32\left(26-41\alpha+64\alpha^2-45\alpha^3+3\alpha^4\right)\beta^3
\notag\\ &\quad
+4\left(104+286\alpha-438\alpha^2+322\alpha^3-204\alpha^4\right)\beta^4
\notag\\ &\quad
-4\alpha\left(105-96\alpha+33\alpha^2-28\alpha^3\right)\beta^5
+2\alpha\left(41+102\alpha-163\alpha^2+34\alpha^3\right)\beta^6
\notag\\ &\quad
-8\alpha^2\left(16-20\alpha+5\alpha^2\right)\beta^7
+\alpha^2\left(15-18\alpha+4\alpha^2\right)\beta^8,
\notag\\ 
C_{14}&
=64(1-\alpha)^2\big[1-(3+2\alpha)\beta-(8-18\alpha)\beta^2
+2\left(4-2\alpha-7\alpha^2\right)\beta^3
\notag\\ &\quad
-\alpha(8-13\alpha)\beta^4-\alpha^2\beta^5\big],
\notag\\ 
C_{12}&
=-32(1-\alpha)\big[2-\alpha-\left(6+3\alpha-2\alpha^2\right)\beta
-\left(16-58\alpha+21\alpha^2\right)\beta^2
\notag\\ &\quad
+\left(14-20\alpha-48\alpha^2+19\alpha^3\right)\beta^3
+\left(14-53\alpha+78\alpha^2-4\alpha^3\right)\beta^4
\notag\\ &\quad
-\left(4-3\alpha-16\alpha^2+36\alpha^3\right)\beta^5
+\alpha\left(4-17\alpha+20\alpha^2\right)\beta^6
+\alpha^2(2-3\alpha)\beta^7\big], 
\notag\\ 
C_{10}&
=4\big[4(1-\alpha)^2\beta^2+\alpha(1-\beta )^4\big]
\big[\alpha-\left(4-\alpha+2\alpha^2\right)\beta
+\left(28-30 \alpha +12 \alpha^2\right)\beta^2
\notag\\ &\quad
-\left(8+22 \alpha -20\alpha^2\right) \beta^3
+\alpha(21-16\alpha)\beta^4-\alpha(3-2\alpha)\beta^5\big],
\notag\\ 
C_{06}&
=256(1-\alpha)^3(1-\beta)\beta(1-\alpha\beta),
\notag\\ 
C_{04}&
=16 (1-\alpha)^2 \big[1-(26-8 \alpha ) \beta 
+\left(41+30\alpha -8\alpha^2\right) \beta^2
-4 \left(1+21\alpha +\alpha^2\right) \beta^3
\notag\\ &\quad
-\left(8-30 \alpha -41 \alpha ^2\right) \beta^4
+2 (4-13 \alpha ) \alpha\beta ^5+\alpha^2 \beta^6\big],
\notag\\ 
C_{02}&
=-8(1-\alpha)(1-\beta)\big[2-\alpha 
-\left(18-7 \alpha +2 \alpha ^2\right) \beta 
+\left(32+16 \alpha +\alpha ^2+2 \alpha ^3\right) \beta^2
\notag\\ &\quad
+\left(24-138\alpha+29\alpha^2-10\alpha^3\right)\beta^3
+\left(10-29\alpha+138\alpha^2-24\alpha^3\right)\beta^4
\notag\\ &\quad
-\left(2+\alpha +16
\alpha^2+32\alpha^3\right)\beta^5
+\alpha\left(2-7\alpha+18\alpha^2\right)\beta^6
+\alpha^2(1-2\alpha)\beta^7\big],
\notag\\ 
C_{00}&
=\left(1-6\beta+\beta^2\right) 
\left[4(1-\alpha)^2\beta^2+\alpha(1-\beta)^4\right]^2.
\end{align}%
\endgroup
Note that the coefficients are polynomials in $\alpha$ 
of the degree at most 4, and in $\beta$ they are all, but $C_{00}$, 
of the degree at most 8; the latter is of the degree 10.

In the limit $\beta\to1$, that is $R\to\Rc$, the arctic curve
factorizes onto two straight lines $Z_1=1$, the usual Arctic ellipse
\eqref{AEZZ}, as expected, and the point $(Z_1,Z_2)=(1,0)$ belonging
to the Arctic ellipse:
\begin{equation}
A(z_1,z_2)\Big|_{\beta=1}=
64(1-\alpha)^6\alpha^6
\left(Z_1-1\right)^2 \left(Z_1^2+Z_2^2-1\right) \left[(Z_1-1)^2+Z_2^2\right].
\end{equation}

In the limit $\beta\to0$, that is $R\to 1$, the arctic curve
factorizes onto two straight lines $Z_1=-1$, and two Arctic ellipses
of radii $1/2$:
\begin{multline}
A(z_1,z_2)\Big|_{\beta=0}=
16(1-\alpha)^4\alpha^6 (Z_1+1)^2
\left[\left(Z_2-\frac{1}{2\sqrt{1-\alpha}}\right)^2+Z_1^2-\frac{1}{4}\right]
\\\times
\left[\left(Z_2+\frac{1}{2\sqrt{1-\alpha}}\right)^2+Z_1^2-\frac{1}{4}\right],
\end{multline}
as expected.


\bibliography{ffac_bib}

\begin{bibdiv}
\begin{biblist}

\bib{KZj-00}{article}{
      author={Korepin, V.~E.},
      author={Zinn-Justin, P.},
       title={Thermodynamic limit of the six-vertex model with domain wall
  boundary conditions},
        date={2000},
     journal={J. Phys. A},
      volume={33},
       pages={7053\ndash 7066},
      eprint={cond-mat/0004250},
}

\bib{Zj-00}{article}{
      author={Zinn-Justin, P.},
       title={Six-vertex model with domain wall boundary conditions and
  one-matrix model},
        date={2000},
     journal={Phys. Rev. E},
      volume={62},
       pages={3411\ndash 3418},
      eprint={math-ph/0005008},
}

\bib{Zj-02}{article}{
      author={Zinn-Justin, P.},
       title={{The influence of boundary conditions in the six-vertex model}},
        date={2002},
      eprint={cond-mat/0205192},
}

\bib{RP-06}{article}{
      author={Reshetikhin, N.},
      author={Palamarchuk, K.},
       title={{The 6-vertex model with fixed boundary conditions}},
        date={2006},
     journal={PoS},
      volume={Solvay},
       pages={012},
      eprint={1010.5011},
}

\bib{CP-09}{article}{
      author={Colomo, F.},
      author={Pronko, A.~G.},
       title={The arctic curve of the domain-wall six-vertex model},
        date={2010},
     journal={J. Stat. Phys.},
      volume={138},
       pages={662\ndash 700},
      eprint={0907.1264},
}

\bib{BL-13}{book}{
      author={Bleher, P.},
      author={Liechty, K.},
       title={{Random Matrices and the Six-Vertex Model}},
      series={{CRM monographs series}},
   publisher={American Mathematical Society},
     address={Providence, RI},
        date={2013},
      volume={32},
}

\bib{RS-15}{article}{
      author={Reshetikhin, N.},
      author={Sridhar, A.},
       title={{Integrability of limit shapes of the six-vertex model}},
        date={2017},
     journal={Commun. Math. Phys.},
      volume={356},
       pages={535\ndash 563},
      eprint={1510.01053},
}

\bib{RS-16}{article}{
      author={Reshetikhin, N.},
      author={Sridhar, A.},
       title={{Limit shapes of the stochastic six-vertex model}},
        date={2016},
      eprint={1609.01756},
}

\bib{ADSV-16}{article}{
      author={Allegra, N.},
      author={Dubail, J.},
      author={St{\'e}phan, J.-M.},
      author={Viti, J.},
       title={Inhomogeneous field theory inside the arctic circle},
        date={2016},
     journal={J. Stat. Mech.},
      volume={2016},
       pages={053108},
      eprint={1512.02872},
}

\bib{BCG-16}{article}{
      author={Borodin, A.},
      author={Corwin, I.},
      author={Gorin, V.},
       title={Stochastic six-vertex model},
        date={2016},
     journal={Duke Math. J.},
      volume={165},
       pages={563\ndash 624},
      eprint={1407.6729},
}

\bib{D-16}{article}{
      author={Dimitrov, E.},
       title={Six-vertex models and the {GUE}-corners process},
        date={2018},
     journal={Int. Math. Res. Notices},
       pages={in press},
      eprint={1610.06893},
}

\bib{GBDJ-18}{article}{
      author={Granet, A.},
      author={Budzynzki, L.},
      author={Dubail, J.},
      author={Jacobsen, J.L.},
       title={{Inhomogeneous Gaussian free field inside the interacting Arctic
  curve}},
      eprint={1807.07927},
}

\bib{JPS-98}{article}{
      author={Jockush, W.},
      author={Propp, J.},
      author={Shor, P.},
       title={{Random domino tilings and the arctic circle theorem}},
      eprint={math/9801068},
}

\bib{CLP-98}{article}{
      author={Cohn, H.},
      author={Larsen, M.},
      author={Propp, J.},
       title={{The shape of a typical boxed plane partition}},
        date={1998},
     journal={New York J. Math.},
      volume={4},
       pages={137\ndash 165},
      eprint={math/9801059},
}

\bib{CK-01}{article}{
      author={Cerf, R.},
      author={Kenyon, R.},
       title={{The low-temperature expansion of the Wulff crystal in the $3D$
  Ising model}},
        date={2001},
     journal={Comm. Math. Phys.},
      volume={222},
       pages={147\ndash 179},
}

\bib{OR-01}{article}{
      author={Okounkov, A.},
      author={Reshetikhin, N.},
       title={{Correlation function of Schur process with application to local
  geometry of a random $3$-dimensional Young diagram}},
        date={2003},
     journal={J. Amer. Math. Soc.},
      volume={16},
       pages={581\ndash 603},
      eprint={math/0107056},
}

\bib{FS-03}{article}{
      author={Ferrari, P.~L.},
      author={Spohn, H.},
       title={{Step fluctuations for a faceted crystal}},
        date={2003},
     journal={J. Stat. Phys.},
      volume={113},
       pages={1\ndash 46},
      eprint={cond-mat/0212456},
}

\bib{KO-06}{article}{
      author={Kenyon, R.},
      author={Okounkov, A.},
       title={{Planar dimers and Harnack curves}},
        date={2006},
     journal={Duke Math. J.},
      volume={131},
       pages={499\ndash 524},
      eprint={math-ph/0311062},
}

\bib{KOS-06}{article}{
      author={Kenyon, R.},
      author={Okounkov, A.},
      author={Sheffield, S.},
       title={Dimers and amoebae},
        date={2006},
     journal={Ann. of Math.},
      volume={163},
       pages={1019\ndash 1056},
      eprint={math-ph/0311005},
}

\bib{KO-07}{article}{
      author={Kenyon, R.},
      author={Okounkov, A.},
       title={{Limit shapes and the complex Burgers equation}},
        date={2007},
     journal={Acta Math.},
      volume={199},
       pages={263\ndash 302},
      eprint={math-ph/0507007},
}

\bib{PS-05}{article}{
      author={Petersen, T.~K.},
      author={Speyer, D.},
       title={An arctic circle theorem for groves},
        date={2005},
     journal={J. Comb. Theory. Series A},
      volume={111},
       pages={137\ndash 164},
      eprint={math/0407171},
}

\bib{PR-07}{article}{
      author={Pittel, B.},
      author={Romik, D.},
       title={Limit shapes for random square {Y}oung tableaux},
        date={2007},
     journal={Adv. Appl. Math},
      volume={38},
       pages={164\ndash 209},
      eprint={math.PR/0405190},
}

\bib{DfSg-14}{article}{
      author={Francesco, P.~Di},
      author={Soto-Garrido, R.},
       title={{Arctic curves of the octahedron equation}},
        date={2014},
     journal={J. Phys. A: Math. Theor.},
      volume={47},
       pages={285204},
      eprint={1402.4493},
}

\bib{Pet-14}{article}{
      author={Petrov, L.},
       title={Asymptotics of random lozenge tilings via {G}elfand--{T}setlin
  schemes},
        date={2014},
     journal={Prob. Theor. and Rel. Fields},
      volume={160},
       pages={429\ndash 487},
      eprint={1202.3901},
}

\bib{RSn-16}{article}{
      author={Romik, D.},
      author={{\'S}niady, P.},
       title={Limit shapes of bumping routes in the {Robinson-Schensted}
  correspondence},
        date={2016},
     journal={Random Struct. Algor.},
      volume={48},
       pages={171\ndash 182},
      eprint={1304.7589},
}

\bib{BBCCR-17}{article}{
      author={Boutillier, C.},
      author={Bouttier, J.},
      author={Chapuy, G.},
      author={Corteel, S.},
      author={Ramassamy, S.},
       title={Dimers on rail yard graphs},
        date={2017},
     journal={Ann. Inst. Henri Poincar\'e Comb. Phys. Interact.},
      volume={4},
       pages={479\ndash 539},
      eprint={1504.05176},
}

\bib{BK-18}{article}{
      author={Bufetov, A.},
      author={Knizel, A.},
       title={Asymptotics of random domino tilings of rectangular {A}ztec
  diamonds},
        date={2018},
     journal={Ann. Inst. H. Poincaré Probab. Statist.},
      volume={54},
       pages={1250\ndash 1290},
      eprint={1604.01491},
}

\bib{DfL-18}{article}{
      author={Francesco, P.~Di},
      author={Lapa, M.F.},
       title={{Arctic curves in path models from the tangent method}},
        date={2018},
     journal={J. Phys. A: Math. Theor.},
      volume={51},
       pages={155202},
      eprint={1711.03182},
}

\bib{DfG-18}{article}{
      author={Francesco, P.~Di},
      author={Guitter, E.},
       title={Arctic curves for paths with arbitrary starting points: a tangent
  method approach},
        date={2018},
     journal={J. Phys. A: Math. Theor.},
      eprint={1803.11463},
        note={In press},
}

\bib{S-17}{article}{
      author={St{\'e}phan, J.-M.},
       title={{Return probability after a quantum quench from a domain wall
  initial state in the spin-1/2 XXZ chain}},
        date={2017},
     journal={J. Stat. Mech. Theory Exp.},
      volume={2017},
       pages={103108},
      eprint={1707.06625},
}

\bib{CDlV-18}{article}{
      author={Collura, M.},
      author={Luca, A.~De},
      author={Viti, J.},
       title={{Analytic solution of the domain wall nonequilibrium stationary
  state}},
        date={2018},
     journal={Phys. Rev. B},
      volume={97},
       pages={081111},
      eprint={1707.06218},
}

\bib{C-17}{article}{
      author={Cugliandolo, L.},
       title={Artificial spin-ice and vertex models},
        date={2017},
     journal={J. Stat. Phys.},
      volume={167},
       pages={499\ndash 514},
      eprint={1701.02283},
}

\bib{K-82}{article}{
      author={Korepin, V.~E.},
       title={{Calculations of norms of Bethe wave functions}},
        date={1982},
     journal={Commun. Math. Phys.},
      volume={86},
       pages={391\ndash 418},
}

\bib{CP-08}{article}{
      author={Colomo, F.},
      author={Pronko, A.~G.},
       title={The limit shape of large alternating-sign matrices},
        date={2010},
     journal={SIAM J. Discrete Math.},
      volume={24},
       pages={1558\ndash 1571},
      eprint={0803.2697},
}

\bib{CPZj-10}{article}{
      author={Colomo, F.},
      author={Pronko, A.~G.},
      author={Zinn-Justin, P.},
       title={The arctic curve of the domain-wall six-vertex model in its
  anti-ferroelectric regime},
        date={2010},
     journal={J. Stat. Mech. Theory Exp.},
       pages={L03002},
      eprint={1001.2189},
}

\bib{CS-16}{article}{
      author={Colomo, F.},
      author={Sportiello, A.},
       title={Arctic curves of the six-vertex model on generic domains: the
  tangent method},
        date={2016},
     journal={J. Stat. Phys.},
      volume={164},
       pages={1488\ndash 1523},
      eprint={1605.01388},
}

\bib{CP-07b}{article}{
      author={Colomo, F.},
      author={Pronko, A.~G.},
       title={Emptiness formation probability in the domain-wall six-vertex
  model},
        date={2008},
     journal={Nucl. Phys. B},
      volume={798},
       pages={340\ndash 362},
      eprint={0712.1524},
}

\bib{J-00}{article}{
      author={Johansson, K.},
       title={{Shape fluctuations and random matrices}},
        date={2000},
     journal={Commun. Math. Phys.},
      volume={209},
       pages={437\ndash 476},
      eprint={math/9903134},
}

\bib{P-13}{article}{
      author={Pronko, A.~G.},
       title={On the emptiness formation probability in the free-fermion
  six-vertex model with domain wall boundary conditions},
        date={2013},
     journal={J. Math. Sci. (N. Y.)},
      volume={192},
       pages={101\ndash 116},
}

\bib{CP-13}{article}{
      author={Colomo, F.},
      author={Pronko, A.~G.},
       title={Third-order phase transition in random tilings},
        date={2013},
     journal={Phys. Rev. E},
      volume={88},
       pages={042125},
      eprint={1306.6207},
}

\bib{CP-15}{article}{
      author={Colomo, F.},
      author={Pronko, A.~G.},
       title={Thermodynamics of the six-vertex model in an {L}-shaped domain},
        date={2015},
     journal={Comm. Math. Phys.},
      volume={339},
       pages={699\ndash 728},
      eprint={1501.03135},
}

\bib{CPS-16}{article}{
      author={Colomo, F.},
      author={Pronko, A.~G.},
      author={Sportiello, A.},
       title={Generalized emptiness formation probability in the six-vertex
  model},
        date={2016},
     journal={J. Phys. A: Math. Theor.},
      volume={49},
       pages={415203},
      eprint={1605.01700},
}

\bib{EKLP-92}{article}{
      author={Elkies, N.},
      author={Kuperberg, G.},
      author={Larsen, M.},
      author={Propp, J.},
       title={{Alternating-sign matrices and domino tilings}},
        date={1992},
     journal={J. Algebraic Combin.},
      volume={1},
       pages={111\ndash 132; 219\ndash 234},
      eprint={math/9201305},
}

\bib{P-03}{article}{
      author={Propp, J.},
       title={Generalized domino-shuffling},
        date={2003},
     journal={Theoret. Computer Sci.},
      volume={303},
       pages={267\ndash 301},
      eprint={math/0111034},
}

\bib{BKMM-07}{book}{
      author={Baik, J.},
      author={Kriecherbauer, T.},
      author={McLaughlin, K. T.-R.},
      author={Miller, P.~D.},
       title={{Discrete orthogonal polynomials: Asymptotics and applications}},
      series={Ann. of Math. Stud.},
   publisher={Princeton University Press},
     address={Princeton, NJ},
        date={2007},
      volume={164},
}

\bib{DK-93}{article}{
      author={Douglas, M.R.},
      author={Kazakov, V.A.},
       title={{Large $N$ phase transition in continuum QCD$_2$}},
        date={1993},
     journal={Phys. Lett. B},
      volume={319},
       pages={219\ndash 230},
      eprint={hep-th/9305047},
}

\bib{ZJ-98}{article}{
      author={Zinn-Justin, P.},
       title={{Universality of correlation functions of Hermitian random
  matrices in an external field}},
        date={1998},
     journal={Comm. Math. Phys.},
      volume={194},
       pages={631\ndash 650},
      eprint={cond-mat/9705044},
}

\end{biblist}
\end{bibdiv}

\end{document}